\documentclass[onecolumn,11pt,oneside,draftclsnofoot]{IEEEtran}%

\usepackage{amsbsy}
\usepackage{floatflt} 

\usepackage{amsmath}
\usepackage{amssymb}
\usepackage{times}
\usepackage{graphicx}
\usepackage{xspace}
\usepackage{paralist} 
\usepackage{setspace} 
\usepackage{xypic}
\xyoption{curve}
\usepackage{latexsym}
\usepackage{theorem}
\usepackage{ifthen}
\usepackage{subfigure}
\usepackage{turnstile}

\usepackage{booktabs}
{\theoremheaderfont{\it} \theorembodyfont{\rmfamily}
\newtheorem{theorem}{Theorem}
\newtheorem{lemma}[theorem]{Lemma}
\newtheorem{definition}[theorem]{Definition}

\newtheorem{proposition}[theorem]{Proposition}
\newtheorem{remark}[theorem]{Remark}

}

\newcommand{\avg}{\mbox{\scriptsize{avg}}}

\title{Convergence Rate Analysis of Distributed Gossip (Linear Parameter) Estimation: Fundamental Limits and Tradeoffs}
\author{Soummya Kar and Jos\'e M.~F.~Moura$^{*}$
\thanks{The first author is with the Dep.~Electrical Engineering, Princeton University, Princeton, NJ. This work was performed while the first author was with the Dep.~Electrical and Computer Engineering, Carnegie Mellon University. The second author is with the Dep.~Electrical and Computer Engineering, Carnegie Mellon University, Pittsburgh, PA 15213, USA (e-mail:
soummyak@andrew.cmu.edu, moura@ece.cmu.edu, ph: (412)268-6341, fax: (412)268-3890.)}
\thanks{Work partially supported by AFOSR grant~\#~FA95501010291; and by NSF grant~\#~CCF1011903.}}
\begin{document}
\maketitle\thispagestyle{empty}\maketitle

\begin{abstract}
The paper considers gossip distributed estimation of a (static) distributed random field (a.k.a., large scale unknown parameter vector) observed by sparsely interconnected sensors, each of which only observes a small fraction of the field. We consider linear distributed estimators whose structure combines the information \emph{flow} among sensors (the \emph{consensus} term resulting from the local gossiping exchange among sensors when they are able to communicate) and the information \emph{gathering} measured by the sensors (the \emph{sensing} or \emph{innovations} term.) This leads to mixed time scale algorithms--one time scale associated with the consensus and the other with the innovations. The paper establishes a distributed observability condition (global observability plus mean connectedness) under which the distributed estimates are consistent and asymptotically normal. We introduce the distributed notion equivalent to the (centralized) Fisher information rate, which is a bound on the mean square error reduction rate of any distributed estimator; we show that under the appropriate modeling and structural network communication conditions (gossip protocol) the distributed gossip estimator attains this distributed Fisher information rate, asymptotically achieving the performance of the optimal centralized estimator. Finally, we study the behavior of the distributed gossip estimator when the measurements fade (noise variance grows) with time; in particular, we consider the maximum rate at which the noise variance can grow and still the distributed estimator being consistent, by showing that, as long as the centralized estimator is consistent, the distributed estimator remains consistent.
\end{abstract}
\hspace{.43cm}\textbf{Keywords:}
Distributed estimation, gossip, random networks, sensor networks, link failures, switching topology

\newpage
\section{Introduction}
\label{introduction}

\subsection{Motivation}
\label{mot}
We consider distributed (or decentralized) estimation of a random field where observations are collected by possibly a large number of sparsely internetworked sensors. The network operates under the gossip random protocol and may be subject to random infrastructure failures (communication channels may fail intermittently.) There is no fusion-center and the estimation is performed locally at each sensor with inter-sensor message exchanges occurring at random times. Because the random field of interest is distributed, each sensor can only observe a part of the field, and no sensor can in isolation obtain a reasonable estimate of the entire field. This paper studies the conditions under which the distributed algorithms operating under the random intermittent conditions (gossip and link failures) that we consider can achieve (asymptotically) performance that is equivalent to the estimation performance of centralized optimal algorithms.
 To be more concrete and as an abstraction of the environment\footnote{The term environment or field has a generic usage here. It may correspond to sensors deployed over a domain of interest like a temperature surface, or, a networked physical system instrumented with sensors. Typical examples of the latter include cyberphysical systems like the power grid, and networked control systems (NCS), where a network of distributed actuators are equipped with sensors.}, we model it by a static vector parameter, whose dimension, $M$, can be arbitrarily large. Each sensor's observations, say for sensor~$n$, are  $M_{n}$ dimensional noisy measurements of a \emph{part} of the (static random) field, where $M_{n}\ll M$. We assume that the sensing rate, i.e.,  rate of receiving observations at each sensor, is comparable to the communication rate among sensors, so that sensors update their estimate at time index~$i$ by fusing appropriately their current estimate with the observation (innovation) at $i$ and the estimates at~$i$ received from those sensors with which it successfully gossips at $i$. Because of the communication intermittency, the distributed estimators that we consider  exhibit mixed time scales: one associated with the \emph{consensus}, i.e., mixing estimation updating resulting from receiving the estimates from the neighbors; and the other associated with the \emph{sensing} or estimation updating from the \emph{innovations}. In this paper, we consider a general class of \emph{linear} distributed gossip networked estimators and study the conditions under which they exhibit the same estimation error convergence rate as a centralized linear field estimator. Nonlinear distributed estimators and distributed estimation of time varying random fields under the gossip protocol are considered elsewhere, \cite{KarMouraRamanan-Est} and~\cite{Gossip-Kalman}, respectively.

We discuss the major challenges in gossip distributed estimation and highlight the key contributions of the paper:
\begin{itemize}
\item \textbf{Infrastructure failures and gossip communication}: The inter-sensor communication may be bandwidth and power constrained and subject to random environmental conditions. For example, the sensors may share a common wireless medium and, due to competing objectives, the inter-sensor transmissions may be scheduled by the underlying MAC (Medium Access Control) layer to occur at random times; in fact, in many situations of interest, the exact medium access (MAC) protocol (randomized) is not known or determined \`apriori, the inter-sensor communications is asynchronous, and random data packet dropouts may occur.
\item \textbf{Distributed observability}: It is well known that centralized estimation requires observability conditions to be satisfied for the estimation task to be \emph{successful}\footnote{Successful means the estimate sequence generated over time possesses desirable properties like consistency, asymptotic normality etc.}. As we will see, formulating a satisfactory notion of \emph{distributed observability} is not trivial. A difficulty stems from the distributed nature of the \emph{information}, i.e., sensors observe only a portion of the field of interest. The incorporation of estimate fusion among the sensor nodes (\emph{consensus}) together with local innovation updates suggest that distributed observability should be not only a function of the sensor observations, but closely tied to the structural properties of the communication network governing the information flow. These conditions are sensitive to the pattern of information dissemination in the network and depends on the level of node cooperation, for example, gossiping.
    We present minimal conditions for distributed observability, namely, for example, in the case of full cooperation (each node exchanges its entire estimate with its neighbors), we show that \emph{global observability}\footnote{Global observability corresponds to the centralized setting, where an estimator has access to the observations of all sensors at all times. The assumption of global observability does not mean that each sensor is observable; rather, that if there was a centralized estimator with simultaneous access to all the sensor measurements, this centralized estimator would be observable.} and mean connectedness of the time varying communication graph are sufficient to ensure \emph{consistent} parameter estimates at \emph{each} sensor.
%
\item  \textbf{Distributed versus optimal centralized estimation}: We show that under reasonable assumptions, the gossip distributed estimators we develop, like the centralized optimal estimator, lead to consistent parameter estimates at each sensor. The natural question of interest is to compare the rate of convergence of these schemes to the true parameter value. We adopt asymptotic normality and the associated asymptotic variance as the metric for comparing different estimators. It is known from the theory of recursive estimation (centralized), that the optimum centralized estimator (under reasonable assumptions) achieves asymptotic variance equal to the Fisher information rate. In this paper, we formalize a notion of distributed Fisher information rate, i.e., a lower bound on the asymptotic variance of all distributed schemes and also investigate the existence of optimal distributed estimators achieving this lower bound. It turns out that, if the inter-sensor communication is noisy or quantized, the asymptotic variance of distributed estimators is always higher than their centralized counterpart. On the other hand, a remarkable asymptotic time scale separation phenomenon shows that, in the absence of channel noise or quantization (but presence of random link failures and gossip,) there exist distributed estimation schemes whose asymptotic variance equals the centralized Fisher information rate under pragmatic conditions. In particular, it is shown that, in a Gaussian environment, a distributed estimator is equivalent to a centralized one in terms of asymptotic variance, and, more generally, equivalent to the best linear centralized estimator. This is significant, as it shows that, under reasonable assumptions, a distributed gossip estimator is as good as a centralized one, the latter having access to all sensor observations at all times. We present some intuitive remarks. In a centralized recursive (parameter) estimation scheme, the estimate update rule involves combining the past estimate with the new innovation (observation), the key design parameter being the time varying gain or weight associated to the innovation term. Since, the observations are noisy, for parameter estimation, this weight sequence needs to go to zero for achieving convergence and, in fact, needs to be square summable to constrain the effect of the observation noise. In most cases, assuming independent observations over time, the innovation gains decrease as $1/i$ ($i$ being the iteration or time index) for optimal estimation performance. This means that the estimation uncertainty cannot be reduced at a rate $1/\sqrt{i}$, a consequence of central limit theorem type arguments. Now, consider the distributed scheme. Here, the algorithm design involves two gain sequences, one for the local innovations at each sensor and the other for estimate fusion (consensus) across sensors. To design good performance distributed gossip estimators, the trick is in choosing the fusion or consensus gain properly, so that its effect decays at a slower rate than the innovation gain. In the absence of quantization or channel noise, it is possible to choose the consensus weight sequence such that its squared sum goes to $\infty$, in contrast to the innovation weight sequence whose squared sum needs to be finite. It is shown that this tuning of the different gain sequences leads to an asymptotic time scale separation, the rate of information dissemination dominating the rate of reduction of uncertainty by observation acquisition. This tuning is not possible in the case of quantized or noisy transmissions, as each consensus step introduces noise, preventing proper adjustment of the gain sequences. The analysis approach that we develop is of independent interest and contributes to the theory of mixed time scale stochastic approximation.\footnote{By mixed time scale, we refer to stochastic algorithms where two potentials act in the same update step with different weight or gain sequences. This should not be confused with stochastic algorithms with coupling (see~\cite{Borkar-stochapp}), where a quickly switching parameter influences the relatively slower dynamics of another state, leading to \emph{averaged} dynamics.} Related to our mixed time scale algorithms is the work~\cite{Gelfand-Mitter}, which develops methods to analyze such algorithms in the context of simulated annealing. In~\cite{Gelfand-Mitter} the role of our innovation potential is played by a martingale difference term. However, in our paper, an additional difficulty with respect to~\cite{Gelfand-Mitter} is that the innovation is not a martingale difference process, and so a key step in our analysis is to derive pathwise strong approximation results to characterize the rate at which the innovation process converges to a martingale difference process.
\end{itemize}

\textbf{Brief review of the literature.} We comment on the relevant literature.  An early treatment of distributed stochastic algorithms
appears in~\cite{tsitsiklisbertsekasathans86} (see also~\cite{tsitsiklisphd84,Bertsekas-survey,Kushner-dist}.)
In~\cite{tsitsiklisbertsekasathans86}, almost sure convergence is established for a class of distributed stochastic algorithms
in the context of distributed optimization. This line of work assumes the existence of a fixed time window $T$, such that the union of communication graphs over any interval of length $T$ is connected with probability one. Also, the stochastic noise appears only in the computation of the local gradients that play the role of innovations in our approach. The conditions imposed on the local gradients are rather strong and implicitly assume that the individual processor (sensor in our terminology) dynamics are stable. Some of these conditions are relaxed in~\cite{Kushner-dist}, which derives almost sure convergence and asymptotic normality for a class of constrained and unconstrained parallel and communicating stochastic procedures with perfect communication. On the contrary, the gossip distributed estimators we develop in this paper are general mixed time-scale procedures in generic random environments and provide pathwise strong convergence rates. Our work does not impose local conditions on the innovation processes and develops and infers connective stability based on structural network conditions and global observability and establishes strong invariance results relating network information flow and the effect of local innovations.

More recently, there has been renewed interest in distributed approaches motivated by wireless sensor networks (WSN) applications.
The papers~\cite{Mesbahi-parameter,Giannakis-est,tsp06-K-A-M,Khan-Moura}
study the estimation problem in static networks, where either the
sensors take a single snapshot of the field at the start and then
initiate distributed consensus protocols (or more generally
distributed optimization, as in~\cite{Giannakis-est}) to fuse the
initial estimates, or the observation rate of the sensors is
assumed to be much slower than the inter-sensor communicate rate,
thus permitting a separation of the two time-scales. More relevant to our work
are~\cite{Sayed-LMS,Stankovic-parameter,Giannakis-LMS,Nedic-parameter},
which consider the linear estimation problem in non-random
networks, where the observation and consensus protocols are
incorporated in the same iteration.
In~\cite{Sayed-LMS,Giannakis-LMS}, the distributed linear
estimation problems are treated in the context of distributed
least-mean-square (LMS) filtering, where constant weight sequences
are used to prove mean-square stability of the filter. The use of
non-decaying combining weights
in~\cite{Sayed-LMS,Giannakis-LMS,Nedic-parameter} leads to a
residual error; however, under appropriate assumptions, these
algorithms can be adapted for tracking certain time-varying
parameters. The distributed LMS algorithm
in~\cite{Stankovic-parameter} considers decaying weight
sequences, thereby establishing $\mathcal{L}_{2}$ convergence to
the true parameter value. In contrast to these, our work quantifies
the pathwise information dissemination rate and its relation to the innovation rate by studying general
mixed time-scale procedures. We consider structural conditions based on the network topology and observation pattern
to develop a satisfactory notion of distributed observability and provide fundamental limits on the performance of
distributed schemes.

The key difference between the current paper and the linear algorithm $\mathcal{LU}$  in~\cite{KarMouraRamanan-Est} involves
 the use of different weight sequences for the consensus and the innovation terms, giving to the linear distributed estimators
 here a mixed time scale behavior. On the other hand, in this paper, we assume unquantized transmissions in the distributed gossip
  estimators.
   Another difference  that will be noted below is the incorporation of
  a general matrix gain $K$ into the innovation update. These modifications make the technical analysis of the distributed gossip linear
  estimators in this paper highly non-trivial and very distinct from the analysis of $\mathcal{LU}$ in~\cite{KarMouraRamanan-Est}.

We briefly comment on the organization of the rest of the paper. Section~\ref{notation} sets up notation and preliminary concepts to be used throughout the paper. Section~\ref{probform} formulates the distributed estimation problem, introduces the algorithm $\mathcal{GLU}$ and the assumptions (Section~\ref{sub-GLU}.) Some technical results on the convergence of stochastic recurrences are established in Section~\ref{inter-res}. This section also considers some properties of centralized estimators, with which we compare our distributed scheme. The main results of the paper are stated in Section~\ref{main_res}. Section~\ref{unq-conv} develops convergence properties of the $\mathcal{GLU}$ algorithm, leading to the proofs of the main theorems in Section~\ref{proof_main_res}. Finally, Section~\ref{conclusion} concludes the paper.

\subsection{Notation}
\label{notation}

We denote the $k$-dimensional Euclidean space by
$\mathbb{R}^{k}$. The set of $m\times n$ matrices with real entries is denoted by $\mathbb{R}^{m\times n}$. $\mathbb{S}^{N},\mathbb{S}_{+}^{N},\mathbb{S}_{++}^{N}$ refer to the subsets of
symmetric, positive semidefinite, positive definite matrices in $\mathbb{R}^{N\times N}$ respectively. The $k\times k$ identity matrix is
denoted by $I_{k}$, while $\mathbf{1}_{k},\mathbf{0}_{k}$ denote
respectively the column vector of ones and zeros in
$\mathbb{R}^{k}$. The set of integers is denoted by $\mathbb{T}$, whereas $\mathbb{N}$ stands for the natural numbers.
$\mathbb{T}_{+}$ denotes the set of nonnegative integers and indices the iteration time slots throughout the paper.

Define the rank one $k\times k$
matrix $P_{k}$ by
\begin{equation}
\label{newnot} P_{k}=\frac{1}{k}\mathbf{1}_{k}\mathbf{1}_{k}^{T}
\end{equation}
The only non-zero eigenvalue of~$P_{k}$ is one, and the corresponding normalized eigenvector is
$\left(1/\sqrt{k}\right)\mathbf{1}_{k}$.

The operator
$\left\|\cdot\right\|$ applied to a vector denotes the standard
Euclidean 2-norm, while applied to matrices denotes the induced
2-norm, which is equivalent to the matrix spectral radius for symmetric
matrices.

We assume that the parameter to be estimated belongs to a subset~$\mathcal{U}$ of the Euclidean space $\mathbb{R}^{M}$.
Throughout the paper, the true (but unknown) value of the
parameter is denoted by~$\mathbf{\theta}^{\ast}$. We denote a
canonical element of~$\mathcal{U}$ by~$\mathbf{\theta}$. The
estimate of~$\mathbf{\theta}^{\ast}$ at time~$i$ at sensor~$n$ is
denoted by $\mathbf{x}_{n}(i)\in\mathbb{R}^{M\times 1}$. Without
loss of generality, we assume that the initial estimate,
$\mathbf{x}_{n}(0)$, at time~$0$ at sensor~$n$ is a non-random
quantity.

Throughout, we assume that all the random objects are defined on a common measurable space, $\left(\Omega,\mathcal{F}\right)$. In case the true (but unknown) parameter value is $\mathbf{\theta}^{\ast}$, the probability and expectation operators are denoted by $\mathbb{P}_{\mathbf{\theta}^{\ast}} \left[\cdot\right]$ and $\mathbb{E}_{\mathbf{\theta}^{\ast}}\left[\cdot\right]$,
respectively. When the context is clear, we abuse notation by dropping the subscript. Also, all inequalities involving random variables are to be interpreted a.s.~(almost surely.)


\textbf{Spectral graph theory.} We review elementary concepts from spectral graph theory. For an \emph{undirected} graph $G=(V,E)$, $V=\left[1\cdots N\right]$ is the set of nodes or vertices, $|V|=N$, and~$E$ is the set of edges, $|E|=M$, where $|\cdot|$ is the cardinality. The unordered pair $(n,l)\in E$ if there exists an edge between nodes~$n$ and~$l$. We only consider simple graphs, i.e., graphs devoid of self-loops and multiple edges. A graph is connected if there exists a path\footnote{A path between nodes $n$ and $l$ of length $m$ is a sequence
$(n=i_{0},i_{1},\cdots,i_{m}=l)$ of vertices, such that, $(i_{k},i_{k+1})\in E\:\forall~0\leq k\leq m-1$.}, between each pair of nodes. The neighborhood of node~$n$ is
\begin{equation}
\label{def:omega} \Omega_{n}=\left\{l\in V\,|\,(n,l)\in
E\right\} 
\end{equation}
Node~$n$ has degree $d_{n}=|\Omega_{n}|$ (number of edges with~$n$ as one end point.) The structure of the graph is described by the symmetric $N\times N$ adjacency matrix, $A=\left[A_{nl}\right]$, $A_{nl}=1$, if $(n,l)\in E$, $A_{nl}=0$, otherwise.
 The degree matrix  is the diagonal matrix $D=\mbox{diag}\left(d_{1}\cdots d_{N}\right)$. The graph positive semi-definite Laplacian matrix, $L$,  and its ordered eigenvalues are
\begin{eqnarray}
\label{def_L} L&=&D-A\\
\label{def_L_eig}
0=\lambda_{1}(L)\leq\lambda_{2}(L)\leq&\cdots&\leq\lambda_{N}(L)
\end{eqnarray}
The smallest eigenvalue $\lambda_{1}(l)$ is always equal to zero, with $\left(1/\sqrt{N}\right)\mathbf{1}_{N}$ being the corresponding normalized eigenvector. The multiplicity of the zero eigenvalue equals the number of connected components of the network; for a connected graph, $\lambda_{2}(L)>0$. This second eigenvalue is the algebraic connectivity or the Fiedler value of
the network; see \cite{FanChung,Mohar,SensNets:Bollobas98} for detailed treatment of graphs and their spectral theory.

\textbf{Kronecker product}: Since, we are dealing with vector parameters, most of the matrix manipulations will involve Kronecker products. For example, the Kronecker product of the $N\times N$ matrix~$L$ and $I_{M}$ will be an $NM\times NM$ matrix, denoted by $L\otimes I_{M}$. Denote the $NM\times NM$ matrix $P^{NM}=P_{N}\otimes I_{M}=\frac{1}{N}(\mathbf{1}_{N}\otimes I_{M})(\mathbf{1}_{N}\otimes I_{M})^{T}$. We will deal often with matrices of the form $C=\left[I_{NM} -bL\otimes I_{M}-aI_{NM}-P^{NM}\right]$, $L$ being a graph Laplacian matrix. It follows from the properties of Kronecker products and the matrices $L,P^{NM}$, that the eigenvalues of this matrix~$C$ are $-a$ and $1-b\lambda_{n}(L)-a,\:n\leq i\leq N$, each being repeated~$M$ times.

\section{Problem Formulation}
\label{probform}
Let $\mathbf{\theta}^{\ast}\in\mathbb{R}^{M\times 1}$ be an $M$-dimensional parameter that is to be estimated by a network of~$N$ sensors. We refer to~$\theta$ as a parameter, although it is a vector of~$M$ parameters. Each sensor makes independent observations of noise corrupted linear functions of the parameter. We assume the following observation model for the $n$-th sensor:
\begin{equation}
\label{obsmod}
\mathbf{z}_{n}(i)=\overline{H}_{n}(i)\mathbf{\theta}^{\ast}+\gamma(i)\mathbf{\zeta}_{n}(i)
\end{equation}
where: \begin{inparaenum}[] \item $\left\{\mathbf{z}_{n}(i)\in\mathbb{R}^{M_{n} \times 1}\right\}_{i\geq 0}$ is the independent observation sequence for the
$n$-th sensor; \item $\left\{\mathbf{\zeta}_{n}(i)\right\}_{i\geq 0}$ is a zero-mean i.i.d.~noise sequence of bounded variance.
\end{inparaenum}
 For most practical sensor network applications, each sensor observes only a subset of~$M_n$ of the components of~$\theta$, with $M_{n}\ll M$. Under such a situation, in isolation, each sensor can estimate at most only a part of the parameter. However, if the sensor network is connected in the mean sense (see assumption (A.3)), and under appropriate observability conditions, we will show that it is possible for each sensor to get a consistent estimate of the parameter~$\mathbf{\theta}^{\ast}$ by means of local inter-sensor communication.

We formalize the assumptions on global observability, fading signal characteristics and network connectivity:

\begin{itemize}
\item{\textbf{(A.1)}}\textbf{Observation Noise}: Recall the
observation model in eqn.~(\ref{obsmod}). We assume that the
process,
$\left\{\mathbf{\zeta}(i)=\left[\mathbf{\zeta}^{T}_{1}(i),
\cdots,\mathbf{\zeta}^{T}_{N}(i)\right]^{T}\right\}_{i\geq 0}$ is
an i.i.d.~zero mean process, with finite second moment. The observation noise process, $\{\gamma(i)\mathbf{\zeta}(i)\}$, then has non-stationary
(in general) characteristics, with variance increasing as $\gamma^{2}(i)$ over time. The non-decreasing sequence $\{\gamma(i)\}$ models the fading characteristics of the parameter (signal) over time. In particular, the regime $\gamma(i)\rightarrow\infty$ corresponds to the SNR decreasing as $1/\gamma^{2}(i)$ over time, whereas, $\gamma(i)=1$ for all $i$ recovers the case of i.i.d. (constant SNR) observations. Also, note that the observation noises at different sensors
may be correlated during a particular iteration, we require only temporal independence. The
spatial correlation of the observation noise makes our model
applicable to practical sensor network problems, for instance, for
distributed target localization, where the observation noise is
generally correlated across sensors.

The following assumption on the growth rate of $\{\gamma(i)\}$ is imposed throughout:

There exists, $0\leq\gamma_{0}<.5$, such that,
\begin{equation}
\label{ass-gamma}
\gamma(i)=(i+1)^{\gamma_{0}},~~\forall i\in\mathbb{T}_{+}
\end{equation}
In other words, we assume that the observation noise variance has sublinear growth. The sublinear growth assumption is not restrictive, and as shown in Remark~\ref{rem:prop-good} is in fact, necessary for centralized estimators to yield consistent estimates of the parameter.


\item{\textbf{(A.2)}}\textbf{Observability:} 
  We require the following global observability condition. The matrix~$G$
\begin{equation}
\label{G} G = \sum_{n=1}^{N}\overline{H}_{n}^{T}\overline{H}_{n}
\end{equation}
is full-rank. This distributed observability extends the
observability condition for a centralized estimator to get a
consistent estimate of the parameter $\mathbf{\theta}^{\ast}$.

\item{\textbf{(A.3)}}\textbf{Random Link Failure:} In digital communications, packets
may be lost at random times. To account for this, we let the links
(or communication channels among sensors) to fail, so that the
edge set and the connectivity graph of the sensor network are time
varying. Accordingly, the sensor network at time~$i$ is modeled as
an undirected graph, $G(i)=(V,E(i))$ and the graph Laplacians as a
sequence of i.i.d.~Laplacian matrices $\left\{L(i)\right\}_{i\geq
0}$. We write
\begin{equation}
\label{Lcond}
L(i) = \overline{L}+\widetilde{L}(i),~\forall i\geq 0
\end{equation}
where the mean $\overline{L} = \mathbb{E}\left[L(i)\right]$. We do
not make any distributional assumptions on the link failure model.
Although the link failures, and so the Laplacians, are independent
at different times, during the same iteration, the link failures
can be spatially dependent, i.e., correlated. This is more general
and subsumes the erasure network model, where the link failures
are independent over space \emph{and} time. Wireless sensor
networks motivate this model since interference among the wireless
communication channels correlates the link failures over space,
while, over time, it is still reasonable to assume that the
channels are memoryless or independent.

Connectedness of the graph is an important issue. We do not
require that the random instantiations~$G(i)$ of the graph be
connected; in fact, it is possible to have all these
instantiations to be disconnected. We only require that the graph
stays connected on \emph{average}. This is captured by requiring
that $\lambda_{2} \left(\overline{L}\right)>0$, enabling us to
capture a broad class of asynchronous communication models; for
example, the random asynchronous gossip protocol analyzed
in~\cite{Boyd-Gossip} satisfies $\lambda_{2}\left(\overline{L}
\right)>0$ and hence falls under this framework.

\item{\textbf{(A.4)}}\textbf{Independence Assumptions}: The
sequences $\left\{L(i)\right\}_{i\in\mathbb{T}_{+}}$ and $\left\{\mathbf{\zeta}(i) \right\}_{i\in\mathbb{T}_{+}}$ are mutually independent.

\end{itemize}


In Section~\ref{sub-GLU}, we present the algorithm $\mathcal{GLU}$ for distributed parameter estimation with the linear observation model~(\ref{obsmod}). Starting from some initial deterministic estimate of the parameters (the initial states may be random, we assume deterministic for notational simplicity), $\mathbf{x}_{n}(0)\in \mathbb{R}^{M\times 1}$, each sensor generates by a distributed iterative algorithm a sequence of estimates, $\left\{ \mathbf{x}_{n}(i)\right\}_{i\geq 0}$. The parameter estimate $\mathbf{x}_{n}(i+1)$ at the $n$-th sensor at time~$i+1$ is a function of: \begin{inparaenum}[] \item its previous estimate; \item the communicated estimates at time~$i$ of its neighboring sensors; and \item the new observation $\mathbf{z}_{n}(i)$.
\end{inparaenum}


\subsection{Algorithm $\mathcal{GLU}$}
\label{sub-GLU}

\textbf{Algorithm $\mathcal{GLU}$}: Consider the parameter estimation problem with linear observation model (assumptions~\textbf{(A.1)-(A.2)}). Let $\mathbf{x}(0)=[\mathbf{x}_{1}(0)^{T},\cdots,\mathbf{x}_{N}(0)^{T}]^{T}$ be the initial estimates of $\mathbf{\theta}^{\ast}$ at the sensors. The $\mathcal{GLU}$ algorithm updates the estimate $\mathbf{x}_{n}(i)$ at sensor $n$ according to the following:
\begin{equation}
\label{unq-GLU1}
\mathbf{x}_{n}(i+1)=\mathbf{x}_{n}(i)-\beta(i)\sum_{l\in\Omega_{n}(i)}(\mathbf{x}_{n}(i)-\mathbf{x}_{l}(i))+\alpha(i)K\overline{H}_{n}^{T}\left(\mathbf{z}_{n}(i)-\overline{H}_{n}\mathbf{x}_{n}(i)\right)
\end{equation}
The key difference between the above scheme and the $\mathcal{LU}$ in~\cite{KarMouraRamanan-Est} involves the use of different weight sequences for the consensus and the innovation terms, giving the former a mixed time scale behavior. On the other hand, we assume unquantized transmissions in $\mathcal{GLU}$. Another difference is the incorporation of a general matrix gain $K$ into the innovation update. These modifications make the technical analysis of $\mathcal{GLU}$ highly non-trivial and different from that of $\mathcal{LU}$, mostly due to the incorporation of mixed time scale dynamics.

In a compact notation, $\mathcal{GLU}$ may be written as:
\begin{equation}
\label{unq-GLU2}
\mathbf{x}(i+1)=\mathbf{x}(i)-\beta(i)\left(L(i)\otimes I_{M}\right)\mathbf{x}(i)+\alpha(i)\left(I_{N}\otimes K\right)\overline{D}_{\overline{H}}\left(\mathbf{z}(i)-D_{\overline{H}}\mathbf{x}(i)\right)
\end{equation}
We refer to the class of distributed recursive estimation algorithms in~\eqref{unq-GLU1}
as $\mathcal{GLU}$. As will be shown, different choices of the weight sequences $\{\alpha(i)\},\{\beta(i)\}$ lead to different convergence characteristics of $\mathcal{GLU}$, hence the usage of the term `class of algorithms'. In the following, we introduce some additional moment requirements and assumptions on the algorithm weight sequences:

\begin{itemize}

\item{\textbf{(A.5)}}\textbf{Moment Condition}: There exists $\varepsilon_{1}>0$, such that,
the following moment exists:
\begin{equation}
\label{unq-GLU700}
\mathbb{E}_{\mathbf{\theta}}\left[\left\|\mathbf{\zeta}(i)\right\|^{2+\varepsilon_{1}}\right]<\infty
\end{equation}
The above implies the existence of a positive function $\kappa_{1}(\cdot)$, such that,
\begin{equation}
\label{unq-GLU5}
\mathbb{E}_{\mathbf{\theta}}\left[\left\|\overline{D}_{\overline{H}}\mathbf{z}(i)-\mathbf{1_{N}}\otimes\left(\left(\frac{1}{N}\mathbf{1}_{N}\otimes I_{M}\right)\overline{D}_{\overline{H}}\mathbf{z}(i)\right)\right\|^{2+\varepsilon_{1}}\right]\leq \gamma^{2+\varepsilon_{1}}(i)\kappa_{1}(\mathbf{\theta})<\infty
\end{equation}
for all $i\in\mathbb{T}_{+}$. We thus assume the existence of slightly greater than quadratic moment of the observation noise process.

\item{\textbf{(A.6)}}\textbf{Weight sequences}: The sequences $\{\alpha(i)\}$ and $\{\beta(i)\}$ are of the form:
\begin{equation}
\label{unq-GLU3}
\alpha(i)=\frac{a}{(i+1)^{\tau_{1}}},\qquad\beta(i)=\frac{b}{(i+1)^{\tau_{2}}}
\end{equation}
where $a,b>0$, $0<\tau_{2}\leq\tau_{1}\leq 1$. In addition, the weights satisfy the following condition:
\begin{equation}
\label{unq-GLU567}
\tau_{1}>\max\left(.5+\gamma_{0}, \tau_{2}+\gamma_{0}+\frac{1}{2+\varepsilon_{1}}\right)
\end{equation}
where $\max(\cdot)$ denotes the maximum of $.5+\gamma_{0}$ and $\tau_{2}+\gamma_{0}+\frac{1}{2+\varepsilon_{1}}$.

The gain matrix $K$ is assumed to be positive definite. To avoid unnecessary technicalities, we also assume that the
matrices $K$ and $G$ commute, so that, $KG$ is symmetric positive definite (see~\cite{Subramanian-Bhagawat}). Recall, $G$ to be the invertible Grammian $\sum_{n=1}^{N}\overline{H}_{n}^{T}\overline{H}_{n}$.

\end{itemize}
\begin{remark}
\label{rem:GLU-ass} We comment on the $\mathcal{GLU}$ assumptions. First, we note that the moment assumption is not restrictive, and most reasonable noise models possess moments of sufficiently high order. Also, it is easy to come up with a choice of algorithm parameters $(\tau_{1},\tau_{2})$ given a $0\leq\gamma<.5$. In fact, any choice of $\tau_{1}>.5+\gamma_{0}$ suffices, as one can choose $\tau_{2}$ satisfying $0<\tau_{2}<\tau_{1}-.5-\gamma_{0}$. That, this choice satisfies assumption (A.6) (\eqref{unq-GLU567}), is due to the fact, that,
$\frac{1}{2+\varepsilon_{1}}<.5$ for any $\varepsilon_{1}>0$. Finally, a note on nomenclature. Often, we will use the term $(\tau_{1},a,\tau_{2},b,K)$-$\mathcal{GLU}$ algorithm to indicate explicitly the $\mathcal{GLU}$ design parameters in force.
\end{remark}

\textbf{Markov.} Consider the filtration, $\left\{\mathcal{F}^{\mathbf{x}}_{i} \right\}_{i\geq 0}$, given by
\begin{equation}
\label{natF}
\mathcal{F}^{\mathbf{x}}_{i}
=\sigma\left(\mathbf{x}(0),\left\{L(j),\mathbf{\zeta}(j)\right\}_{0\leq j<i}\right)
\end{equation}
It then follows that the random objects $L(i),\mathbf{z}(i)$ are independent of
$\mathcal{F}^{\mathbf{x}}_{i}$, rendering $\left\{\mathbf{x}(i),
\mathcal{F}^{\mathbf{x}}_{i}\right\}_{i\in\mathbb{T}_{+}}$ a Markov process.

\subsection{Centralized linear estimators}
\label{lin-cent-est}
The key focus of the paper is to compare the performance achieved by the class of $\mathcal{GLU}$ algorithms to centralized estimation schemes\footnote{A centralized scheme corresponds to a fusion center having access to all sensor observations at all times.}. Specifically, we will restrict this comparison to linear centralized estimators only. To this end, we start by defining \emph{a reasonable} (to be clear soon) class of centralized linear\footnote{Since we deal with linear centralized estimators only, in the following we drop the term linear when referring to centralized estimators.} of the parameter $\mathbf{\theta}$.
\begin{definition}[Centralized linear estimator]
\label{def_centest}
 A centralized linear estimator is a process $\{\mathbf{u}(i)\}_{i\in\mathbb{T}_{+}}$ evolving as
\begin{equation}
\label{uc3:1}
\mathbf{u}(i+1)=\mathbf{u}(i)+\frac{\alpha_{c}(i)}{N}K_{c}\sum_{n=1}^{N}\left(\overline{H}_{n}^{T}\mathbf{z}_{n}(i)-\overline{H}_{n}^{T}\overline{H}_{n}\mathbf{u}(i)\right)
\end{equation}
Here, we assume that the weight sequence $\{\alpha_{c}(i)\}$ is of the form
\begin{equation}
\label{cent-est}
\alpha_{c}(i)=\frac{a_{c}}{(i+1)^{\tau_{c}}}
\end{equation}
for some $a_{c}>0$ and $\tau_{c}\geq 0$.
Also, $K_{c}$ is a positive definite gain matrix that commutes with the Grammian $G$.

A centralized linear estimator is called \emph{good}, if in addition the design parameter satisfies
\begin{equation}
\label{cent-est217}
.5+\gamma_{0}<\tau_{c}\leq 1
\end{equation}
\end{definition}
\begin{remark}
\label{rem:cent}
We comment on the above definition and justify the nomenclature good. Clearly, different choices of the gain matrix $K_{c}$ and the weight sequence $\{\alpha_{c}(i)\}$ would lead to different convergence properties of the estimator $\{\mathbf{u}(i)\}$. As shown in Proposition~\ref{prop-good}, the condition $.5+\gamma_{0}<\tau_{c}\leq 1$ is necessary and sufficient for the estimator $\{\mathbf{u}(i)\}$ to be universally\footnote{By universal consistency of an algorithm, we mean that the algorithm leads to consistent estimates of the parameter $\mathbf{\theta}$ irrespective of the observation noise distribution, as long as the moment assumption (A.5) is satisfied.} consistent from all initial conditions.  In particular, the best linear centralized estimator assumes the form in Definition~\ref{def_centest} (for a specific choice of $K_{c}$ and $\{\alpha_{c}(i)\}$.) Hence, for all purposes, it is sufficient to compare the distributed algorithm $\mathcal{GLU}$ with the class of good centralized estimators defined above. In the following, we will restrict attention to good centralized estimators only, and will often drop the term good when referring to these estimators. Also, similar to the distributed $\mathcal{GLU}$ estimators,  we will use the term $(\tau_{c},a_{c},K_{c})$ centralized estimator to indicate explicitly the design parameters in force.
\end{remark}

Before proceeding to the convergence analysis of $\mathcal{GLU}$ under assumptions~\textbf{(A.1)-(A.6)}, we establish some properties of general stochastic recursions to be used in the sequel.

\section{Some Intermediate Results}
\label{inter-res}

We establish three approximation results to be used later.
The first one (Lemma~\ref{unq-GLU_lemma}) is a stochastic analogue of Lemma 18 in~\cite{KarMouraRamanan-Est}, the second one (Lemma~\ref{unq-prop}) quantifies the pathwise convergence rate in Lemma~\ref{unq-GLU_lemma}. Lemma~\ref{unq-GLU_lm1} is a time-varying mixed time scale version of Lemma 3 in~\cite{KarMouraRamanan-Est}. Finally, we end this section by listing some convergence properties of the centralized estimators (Definition~\ref{def_centest}.)

\begin{lemma}
\label{unq-GLU_lemma}
Consider the scalar time-varying linear system:
\begin{equation}
y(i+1)=(1-r_{1}(i))y(i)+r_{2}(i)
\end{equation}
Here $\{r_{1}(i)\}$ is a sequence of independent random variables, such that, $0\leq r_{1}(i)\leq 1$ a.s. with mean
\begin{equation}
\label{ung-GLU_lemma1}
\overline{r}_{1}(i)=\frac{a_{1}}{(i+1)^{\delta_{1}}}
\end{equation}
and $a_{1}> 0$, $0\leq\delta_{1}\leq 1$. Also, assume $y(0)\geq 0$ and the sequence $\{r_{2}(i)\}$ is given by
\begin{equation}
\label{unq-GLU_lemma2}
r_{2}(i)=\frac{a_{2}}{(i+1)^{\delta_{2}}}
\end{equation}
where $a_{2}>0,\delta_{2}\geq 0$. Then, if $\delta_{1}<\delta_{2}$,
\begin{equation}
\label{unq-GLU_lemma3}
\lim_{i\rightarrow\infty}y(i)=0~\mbox{a.s.}
\end{equation}
\end{lemma}
\begin{proof}
The assumptions imply that the sequence $\{y(i)\}$ is non-negative. Define the process $\{V_{1}(i)\}$ by
\begin{equation}
\label{unq-GLU_lemma4}
V_{1}(i)=y(i)-\sum_{k=0}^{i-1}\left[\left(\prod_{l=k+1}^{i-1}(1-\overline{r}_{1}(l))\right)r_{2}(k)\right]
\end{equation}
Since $\delta_{1}<\delta_{2}$, an application of Lemma 18 in~\cite{KarMouraRamanan-Est} yields
\begin{equation}
\label{unq-GLU_lemma5}
\lim_{i\rightarrow\infty}\sum_{k=0}^{i-1}\left[\left(\prod_{l=k+1}^{i-1}(1-\overline{r}_{1}(l))\right)r_{2}(k)\right]=0
\end{equation}
Hence, in particular, the second term on the R.H.S. is bounded and $\{y(i)\}$ is well defined. Denote by $\{\mathcal{F}^{y}(i)\}$ the natural filtration of the process $\{y(i)\}$ and note that $\{V_{1}(i)\}$ is adapted to this filtration.  Using the fact, that
\begin{equation}
\label{unq-GLU_lemma6}
\sum_{k=0}^{i}\left[\left(\prod_{l=k+1}^{i}(1-\overline{r}_{1}(l))\right)r_{2}(k)\right]=(1-\overline{r}_{1}(i))\left[\sum_{k=0}^{i-1}\left[\left(\prod_{l=k+1}^{i-1}(1-\overline{r}_{1}(l))\right)r_{2}(k)\right]\right]+r_{2}(i)
\end{equation}
we have, by the independence condition,
\begin{eqnarray}
\label{unq-GLU_lemma7}
\mathbb{E}\left[V_{1}(i+1)~|~\mathcal{F}^{y}(i)\right] & = &  \mathbb{E}\left[y(i+1)~|~\mathcal{F}^{y}(i)\right]-\sum_{k=0}^{i}\left[\left(\prod_{l=k+1}^{i}(1-\overline{r}_{1}(l))\right)r_{2}(k)\right]\nonumber \\ & = & (1-\overline{r}_{1}(i))y(i)+r_{2}(i)-\sum_{k=0}^{i}\left[\left(\prod_{l=k+1}^{i}(1-\overline{r}_{1}(l))\right)r_{2}(k)\right]\nonumber \\ & = & (1-\overline{r}_{1}(i))y(i)-\sum_{k=0}^{i-1}\left[\left(\prod_{l=k+1}^{i-1}(1-\overline{r}_{1}(l))\right)r_{2}(k)\right]\nonumber \\ & = &  V_{1}(i)-\overline{r}_{1}(i)y(i)
\end{eqnarray}
The nonnegativity of $\{y(i)\}$ implies
\begin{equation}
\label{unq-GLU_lemma8}
\mathbb{E}\left[V_{1}(i+1)~|~\mathcal{F}^{y}(i)\right]\leq V_{1}(i)
\end{equation}
Hence $\{V_{1}(i)\}$ is a supermartingale. The nonnegativity of $\{y(i)\}$ and the boundedness of the terms $\sum_{k=0}^{i-1}\left[\left(\prod_{l=k+1}^{i-1}(1-\overline{r}_{1}(l))\right)r_{2}(k)\right]$ for all $i$ show that $\{V_{1}(i)\}$ is bounded from below. It then follows that there exists a finite random variable $V_{1}^{\ast}$, such that,
\begin{equation}
\label{unq-GLU_lemma9}
\lim_{i\rightarrow\infty}V_{1}(i)=V_{1}^{\ast}~\mbox{a.s.}
\end{equation}
We then have
\begin{eqnarray}
\label{unq-GLU_lemma10}
\lim_{i\rightarrow\infty}y(i) & = & \lim_{i\rightarrow\infty}V_{1}(i)+\lim_{i\rightarrow\infty}\sum_{k=0}^{i-1}\left[\left(\prod_{l=k+1}^{i-1}(1-\overline{r}_{1}(l))\right)r_{2}(k)\right]\nonumber \\ & = & V_{1}^{\ast}
\end{eqnarray}
Since $y(0)$ is deterministic, the sequence $\{y(i)\}$ is integrable and we have
\begin{equation}
\label{unq-GLU_lemma11}
\mathbb{E}\left[y(i)\right]=\left(\prod_{k=0}^{i}(1-\overline{r}_{1}(i))\right)y(0)+\sum_{k=0}^{i-1}\left[\left(\prod_{l=k+1}^{i-1}(1-\overline{r}_{1}(l))\right)r_{2}(k)\right]
\end{equation}
An application of Lemma 18 in~\cite{KarMouraRamanan-Est} then shows
\begin{equation}
\label{unq-GLU_lemma12}
\lim_{i\rightarrow\infty}\mathbb{E}\left[y(i)\right]=0
\end{equation}
and by Fatou's lemma we conclude $\mathbb{E}\left[V_{1}^{\ast}\right]=0$. Since, $V_{1}^{\ast}$ is nonnegative, being the limit of the nonnegative sequence $\{y(i)\}$, we have
\begin{equation}
\label{unq-GLU_lemma13}
V_{1}^{\ast}=0~\mbox{a.s.}
\end{equation}
and the claim holds.
\end{proof}

We will also use the following result, which characterizes the convergence rate in the above. The proof is somewhat similar to the arguments in Lemma~\ref{unq-GLU_lemma} and we omit it due to space limitations.
\begin{lemma}
\label{unq-prop}
Consider the scalar deterministic time-varying linear system:
\begin{equation}
y(i+1)=(1-r_{1}(i))y(i)+r_{2}(i)
\end{equation}
where the sequences $\{r_{1}(i)\}$ and $\{r_{2}(i)\}$ satisfy the hypothesis of Lemma~\ref{unq-GLU_lemma}.
\begin{itemize}
\item{\textbf{(1)}}
Then, if $\delta_{1}<\delta_{2}$ and $\delta_{1}<1$,
\begin{equation}
\lim_{i\rightarrow\infty}(i+1)^{\delta_{0}}y(i)=0
\end{equation}
for all $0\leq\delta_{0}<\delta_{2}-\delta_{1}$.
\item{\textbf{(2)}} Let $\delta_{1}<\delta_{2}$ and $\delta_{1}=1$. Then the above conclusion holds, if in addition $a_{1}>\delta_{0}$.
\item{\textbf{(3)}} All the above remain valid when $r_{1}(i)$ is random satisfying the conditions of Lemma~\ref{unq-GLU_lemma}.
\end{itemize}
\end{lemma}

\begin{lemma}
\label{unq-GLU_lm1}
Under the stated assumptions, there exists $i_{1}$ sufficiently large and a constant $c_{4}>0$, such that, for $i\geq i_{1}$,
\begin{equation}
\label{unq-GLU_lm12}
\mathbf{y}^{T}\left(\beta(i)\overline{L}\otimes I+\alpha(i)(I_{N}\otimes K)D_{\overline{H}}\right)\mathbf{y}\geq c_{4}\alpha(i)\left\|\mathbf{y}\right\|^{2},~~~\forall\mathbf{y}\in\mathbb{R}^{NM}
\end{equation}
\end{lemma}
\begin{proof}
The key difference from the proof of Lemma 3 in~\cite{KarMouraRamanan-Est} is that, the matrix $\left(\beta(i)\overline{L}\otimes I+\alpha(i)(I_{N}\otimes K)D_{\overline{H}}\right)$ is not symmetric. We first show that the quadratic form
\begin{equation}
\label{exce100}
\mathbf{y}^{T}\left(\frac{\beta(i)}{\alpha(i)}\overline{L}\otimes I+(I_{N}\otimes K)D_{\overline{H}}\right)\mathbf{y}
\end{equation}
is strictly greater than zero for all $\mathbf{y}\in\mathbb{R}^{NM}$ satisfying $\|\mathbf{y}\|=1$ for all sufficiently large $i$. To this end, for such $\mathbf{y}$, consider the decomposition
\begin{equation}
\label{exce101}
\mathbf{y}=\mathbf{y}_{\mathcal{C}}+\mathbf{y}_{\mathcal{C}^{\perp}}
\end{equation}
Define the symmetric matrix $D_{K}$ by
\begin{equation}
\label{exce11}
D_{K}=\frac{1}{2}\left[(I_{N}\otimes K)D_{\overline{H}}\right]+\frac{1}{2}\left[(I_{N}\otimes K)D_{\overline{H}}\right]^{T}
\end{equation}
Noting that
\begin{equation}
\label{exce102}
\mathbf{y}^{T}\left[(I_{N}\otimes K)D_{\overline{H}}\right]\mathbf{y}=\mathbf{y}^{T}D_{K}\mathbf{y}
\end{equation}
we have
\begin{eqnarray}
\label{exce103}
\mathbf{y}^{T}\left(\frac{\beta(i)}{\alpha(i)}\overline{L}\otimes I+(I_{N}\otimes K)D_{\overline{H}}\right)\mathbf{y} & = & \mathbf{y}^{T}\left(\frac{\beta(i)}{\alpha(i)}\overline{L}\otimes I+D_{K}\right)\mathbf{y}\nonumber \\ & = & \mathbf{y}^{T}\left(\frac{\beta(i)}{\alpha(i)}\overline{L}\otimes I\right)\mathbf{y} + \mathbf{y}^{T}D_{K}\mathbf{y}\nonumber \\ & = & \mathbf{y}_{\mathcal{C}^{\perp}}^{T}\left(\frac{\beta(i)}{\alpha(i)}\overline{L}\otimes I\right)\mathbf{y}_{\mathcal{C}^{\perp}}+\mathbf{y}_{\mathcal{C}^{\perp}}^{T}D_{K}\mathbf{y}_{\mathcal{C}^{\perp}}\nonumber \\ & & +2\mathbf{y}_{\mathcal{C}^{\perp}}^{T}D_{K}\mathbf{y}_{\mathcal{C}}+\mathbf{y}_{\mathcal{C}}^{T}D_{K}\mathbf{y}_{\mathcal{C}}\nonumber \\ & \geq & \frac{\beta(i)}{\alpha(i)}\lambda_{2}(\overline{L})\left\|\mathbf{y}_{\mathcal{C}^{\perp}}\right\|^{2}+\mathbf{y}_{\mathcal{C}^{\perp}}^{T}D_{K}\mathbf{y}_{\mathcal{C}^{\perp}}\nonumber \\ & & +2\mathbf{y}_{\mathcal{C}^{\perp}}^{T}D_{K}\mathbf{y}_{\mathcal{C}}+\mathbf{y}_{\mathcal{C}}^{T}D_{K}\mathbf{y}_{\mathcal{C}}
\end{eqnarray}
Now, the symmetricity of $D_{K}$ implies the existence of a constant $c_{15}>0$, large enough, such that,
\begin{eqnarray}
\label{exce104}
\mathbf{y}_{\mathcal{C}^{\perp}}^{T}D_{K}\mathbf{y}_{\mathcal{C}^{\perp}} & \geq & -c_{15}\left\|\mathbf{y}_{\mathcal{C}^{\perp}}\right\|^{2}\\
\mathbf{y}_{\mathcal{C}^{\perp}}^{T}D_{K}\mathbf{y}_{\mathcal{C}} & \geq & -c_{15}\left\|\mathbf{y}_{\mathcal{C}}\right\|\left\|\mathbf{y}_{\mathcal{C}^{\perp}}\right\|
\end{eqnarray}
Also, using the form $\mathbf{y}_{\mathcal{C}}=\mathbf{1}_{N}\otimes\mathbf{a}$, for some $\mathbf{a}\in\mathbb{R}^{M}$, we note that
\begin{eqnarray}
\label{exce105}
\mathbf{y}_{\mathcal{C}}^{T}D_{K}\mathbf{y}_{\mathcal{C}} & = & \mathbf{y}_{\mathcal{C}}^{T}\left[(I_{N}\otimes K)D_{\overline{H}}\right]\mathbf{y}_{\mathcal{C}}\nonumber \\ & = & \sum_{n=1}^{N}\mathbf{a}^{T}K\overline{H}_{n}\mathbf{a}\nonumber \\ & = & \mathbf{a}KG\mathbf{a}\nonumber \\ & \geq & \lambda_{\mbox{\scriptsize{min}}}\left\|\mathbf{a}\right\|^{2}\nonumber \\ & = & \frac{\lambda_{\mbox{\scriptsize{min}}}}{N}\left\|\mathbf{y}_{\mathcal{C}}\right\|^{2}
\end{eqnarray}
where the last but one step uses the fact, that the matrix $KG$ is positive definite, as both $K$ and $G$ are positive definite and they commute. Note, in particular, that $\lambda_{\mbox{\scriptsize{min}}}>0$. Substituting the above in eqn.~(\ref{exce103}), we have
\begin{equation}
\label{exce106}
\mathbf{y}^{T}\left(\frac{\beta(i)}{\alpha(i)}\overline{L}\otimes I+(I_{N}\otimes K)D_{\overline{H}}\right)\mathbf{y} \geq \left(\frac{\beta(i)}{\alpha(i)}\lambda_{2}(\overline{L})-c_{15}\right)\left\|\mathbf{y}_{\mathcal{C}^{\perp}}\right\|^{2}-2c_{15}\left\|\mathbf{y}_{\mathcal{C}}\right\|\left\|\mathbf{y}_{\mathcal{C}^{\perp}}\right\|+\frac{\lambda_{\mbox{\scriptsize{min}}}}{N}\left\|\mathbf{y}_{\mathcal{C}}\right\|^{2}
\end{equation}
Since $\lim_{i\rightarrow\infty}\beta(i)/\alpha(i)=\infty$ ($\tau_{2}<\tau_{1}$), we can choose $i_{1}$ large enough, such that, for $i\geq i_{0}$
\begin{eqnarray}
\label{exce107}
\frac{\beta(i)}{\alpha(i)}\lambda_{2}(\overline{L})-c_{15} & > & 0\\
\frac{\lambda_{\mbox{\scriptsize{min}}}}{N}\left[\frac{\beta(i)}{\alpha(i)}-c_{15}\right] & > & c_{15}^{2}
\end{eqnarray}
We now verify the claim in eqn.~(\ref{exce100}) for $i\geq i_{1}$. Clearly, if $\mathbf{y}_{\mathcal{C}}=\mathbf{0}$, the quadratic form reduces to
\begin{equation}
\label{exce108}
\mathbf{y}^{T}\left(\frac{\beta(i)}{\alpha(i)}\overline{L}\otimes I+(I_{N}\otimes K)D_{\overline{H}}\right)\mathbf{y}\geq \left(\frac{\beta(i)}{\alpha(i)}\lambda_{2}(\overline{L})-c_{15}\right)\left\|\mathbf{y}_{\mathcal{C}^{\perp}}\right\|^{2}=\frac{\beta(i)}{\alpha(i)}\lambda_{2}(\overline{L})-c_{15}>0
\end{equation}
(Note that, the constraint that $\mathbf{y}$ lies on the unit circle forces $\left\|\mathbf{y}_{\mathcal{C}^{\perp}}\right\|$ to be 1, if $\mathbf{y}_{\mathcal{C}}=\mathbf{0}$.)
On the other hand, if $\mathbf{y}_{\mathcal{C}}>0$, we have
\begin{equation}
\label{exce109}
\mathbf{y}^{T}\left(\frac{\beta(i)}{\alpha(i)}\overline{L}\otimes I+(I_{N}\otimes K)D_{\overline{H}}\right)\mathbf{y}\geq \left\|\mathbf{y}_{\mathcal{C}}\right\|^{2}\left[\left(\frac{\beta(i)}{\alpha(i)}\lambda_{2}(\overline{L})-c_{15}\right)\frac{\left\|\mathbf{y}_{\mathcal{C}^{\perp}}\right\|^{2}}{\left\|\mathbf{y}_{\mathcal{C}}\right\|^{2}}-2c_{15}\frac{\left\|\mathbf{y}_{\mathcal{C}^{\perp}}\right\|}{\left\|\mathbf{y}_{\mathcal{C}}\right\|}+\frac{\lambda_{\mbox{\scriptsize{min}}}}{N}\right]
\end{equation}
The term on the R.H.S. is always strictly greater than zero by the discriminant condition of eqn.~(\ref{exce107}).

The assertion in eqn.~(\ref{exce100}) thus holds. Since the quadratic form is a continuous function of $\mathbf{y}$, its positivity on the unit circle implies, there exists $c_{4}>0$, such that,
\begin{equation}
\label{exce110}
\inf_{\|\mathbf{y}\|=1}\mathbf{y}^{T}\left(\frac{\beta(i)}{\alpha(i)}\overline{L}\otimes I+(I_{N}\otimes K)D_{\overline{H}}\right)\mathbf{y}\geq c_{4}>0
\end{equation}
It then follows that, for all $\mathbf{y}\in\mathbb{R}^{NM}$,
\begin{equation}
\label{exce111}
\mathbf{y}^{T}\left(\frac{\beta(i)}{\alpha(i)}\overline{L}\otimes I+(I_{N}\otimes K)D_{\overline{H}}\right)\mathbf{y}\geq c_{4}\left\|\mathbf{y}\right\|^{2}
\end{equation}
and hence
\begin{eqnarray}
\label{exce112}
\mathbf{y}^{T}\left(\beta(i)\overline{L}\otimes I+\alpha(i)(I_{N}\otimes K)D_{\overline{H}}\right)\mathbf{y} & = & \alpha(i)
\mathbf{y}^{T}\left(\frac{\beta(i)}{\alpha(i)}\overline{L}\otimes I+(I_{N}\otimes K)D_{\overline{H}}\right)\mathbf{y}\nonumber \\ & \geq & \alpha(i)c_{4}\left\|\mathbf{y}\right\|^{2}
\end{eqnarray}
for $i\geq i_{1}$.
\end{proof}
Note that, the condition $\lim_{i\rightarrow\infty}\beta(i)/\alpha(i)=\infty$ is required for Lemma~\ref{unq-GLU_lm1}.

The following proposition justifies the nomenclature good in Definition~\ref{def_centest}. In particular, it shows that under assumptions (A.1),(A.2),(A.5), there exists a noise distribution (Gaussian), such that, the centralized scheme is not consistent if $\tau_{c}$ fails to satisfy the requirement~\eqref{cent-est217}.
\begin{proposition}
\label{prop-good}
\begin{itemize}
\item[(1)] Suppose the process $\{\mathbf{\zeta}(i)\}$ is Gaussian. Consider the centralized estimator $\{\mathbf{u}(i)\}$. Then, if $\tau_{c}\leq \gamma_{0}+.5$ or $\tau_{c}>1$, the sequence $\{\mathbf{u}(i)\}$ is not consistent from arbitrary initial condition $\mathbf{u}(0)$.\\
\item[(2)] Let assumptions (A.1),(A.2),(A.5) hold. Then, a good centralized estimator is consistent (universally) from all initial conditions.\\

\item[(3)] Let assumptions (A.1),(A.2),(A.5) hold. Consider a good centralized estimator with design parameters $(\tau_{c},a_{c},K_{c})$. Then, there exists a $(\tau_{1},a,\tau_{2},b,K)$-$\mathcal{GLU}$ estimator, such that, $\tau_{1}=\tau_{c}$, $a=a_{c}$, $K=K_{c}$.
\end{itemize}
\end{proposition}
\begin{remark}
\label{rem:prop-good}
As a consequence of the first assertion, we note that, for a centralized linear estimator to achieve consistency, the parameter $\gamma_{0}$ should be strictly less than .5.
\end{remark}
\begin{proof}
Due to space limitations, we omit the proof which follows from standard properties of stochastic recurrences and approximation (\cite{Nevelson}).

We present an intuitive sketch of the proof of the first assertion. From~\eqref{uc3:1}, we note that, at time $i$, an observation noise is incorporated on the right hand side (R.H.S.) with variance of the order $(i+1)^{2\gamma_{0}-2tau_{c}}$. Clearly, if $\tau_{c}\leq .5+\gamma_{0}$, as $i\rightarrow\infty$ the cumulative noise adds up to $\infty$. For Gaussian noise, this would lead to unboundedness of the estimate sequence $\{\mathbf{u}(i)\}$. This explains the lower bound in the choice of $\tau_{c}$. On the other hand, if $\tau_{c}>1$, the $\{\alpha_{c}\}$ becomes summable and the updates die out quickly. Hence, depending on the initial estimate $\mathbf{u}(0)$, it may not be possible to progress towards $\mathbf{\theta}^{\ast}$. Thus, in general, we need $\tau_{c}\leq 1$.

The second assertion follows from standard stochastic approximation arguments (see, for example~\cite{Nevelson} and Theorem 1 in~\cite{karmoura-randomtopologynoise}.)

The third assertion simply states that there exists a choice of $\tau_{2}$ satisfying assumption (A.6), when $\tau_{1}=\tau_{c}$ and $K=K_{c}$. This is immediate from Remark~\ref{rem:GLU-ass}.
\end{proof}

In the case $\gamma_{0}=1$, i.e., the observation process is stationary (constant SNR), the following property of $\{\mathbf{u}(i)\}$ holds:
\begin{proposition} Suppose $\gamma_{0}=0$ and assumptions (A.1),(A.2),(A.5) hold. Then, in addition to the consistency in Proposition~\ref{prop-good}, we have the following:
\label{prop-centconv}
\begin{itemize}
\item[(1)]
Assume $\tau_{c}=1$, i.e., the weight sequence $\{\alpha_{c}(i)\}$ is of the form
\begin{equation}
\alpha_{c}(i)=\frac{a_{c}}{i+1}
\end{equation}
Then, if $a_{c}>\frac{N}{2\lambda_{\mbox{\scriptsize{min}}}(KG)}$, the normalized sequence $\{1/\sqrt{(i+1)}(\mathbf{u}(i)-\mathbf{\theta}^{\ast})\}$ is asymptotically normal, i.e.,
\begin{equation}
\label{uc3:4}
\sqrt(i+1)\left(\mathbf{u}(i)-\mathbf{\theta}^{\ast}\right)\Longrightarrow\mathcal{N}(\mathbf{0},S_{c}(K))
\end{equation}
where, the asymptotic variance is given by:
\begin{eqnarray}
\label{uc3:5}
S_{c}(K) & = & \frac{a^{2}}{N^{2}}\int_{0}^{\infty}e^{\Sigma_{1}v}S_{1}e^{\Sigma^{T}v}dv\\
\Sigma_{1} & = & -\frac{a}{N}KG+\frac{1}{2}I_{M}\\
S_{1} & = & K\left(\mathbf{1}_{N}\otimes I_{M}\right)^{T}\overline{D}_{\overline{H}}S_{\mathbf{\zeta}}\overline{D}_{\overline{H}}^{T}\left(\mathbf{1}_{N}\otimes I_{M}\right)K^{T}
\end{eqnarray}
\item[(2)] Let the hypothesis of the previous assertion hold and choose $K_{c}=K^{\ast}_{c}=G^{-1}$. Then, the estimator $\{\mathbf{u}(i)\}$ is the best linear centralized estimator in terms of asymptotic variance irrespective of the distribution of the observation noise $\mathbf{\zeta}(i)$. In addition, if the observation noise sequence $\{\mathbf{\zeta}(i)\}$ is Gaussian, $\{\mathbf{u}(i)\}$ as defined above, is the optimum centralized estimator, whose asymptotic variance $S_{c}(K^{\ast})$ equals the centralized Fisher information rate.
\end{itemize}
\end{proposition}
\begin{proof}
The proof of the first assertion is omitted due to space limitations (see~\cite{KarMouraRamanan-Est} for similar arguments.) That, $K_{c}=G^{-1}$ yields the best linear estimator is standard (see, for example,~\cite{Scharf}.)
\end{proof}

\section{Main Results}
\label{main_res}
\begin{theorem}
\label{main1}
Consider a fixed $0\leq \gamma_{0}<.5$. Let assumptions (A.1),(A.2),(A.5) hold.
\begin{itemize}
\item[(1)]
 Consider the $\mathcal{GLU}$ algorithm with design parameters $(\tau_{1},a,\tau_{2},b,K)$ satisfying assumption (A.6). For each sensor $n$, the estimate sequence $\{\mathbf{x}_{n}(i)\}$ generated by the $\mathcal{GLU}$ is a consistent estimator of $\mathbf{\theta}^{\ast}$, i.e.,
\begin{equation}
\label{main1:1}
\mathbb{P}_{\mathbf{\theta}^{\ast}}\left(\lim_{i\rightarrow\infty}\mathbf{x}_{n}(i)=\mathbf{\theta}^{\ast}\right)=1,~~\forall n
\end{equation}
\item[(2)] Consider a centralized estimator $\{\mathbf{u}(i)\}$ corresponding to a given choice of $\{\alpha_{c}\}$ and $K_{c}$.
    Choose $K=K_{c}$, $\tau_{1}=\tau_{c}$ and $\tau_{2}$ satisfying $0<\tau_{2}<\tau_{1}-\gamma_{0}-\frac{1}{2+\varepsilon_{1}}$, such that, assumptions (A.1)-(A.6) hold (such a choice is always possible by Proposition~\ref{prop-good}.) Also, if $\tau_{1}=1$, further assume that the constant $a$ in assumption~\textbf{(A.6)} satisfies
\begin{equation}
\label{main1:2}
a>\frac{N\tau_{0}}{\lambda_{\mbox{\scriptsize{min}}}(KG)}
\end{equation}
For each sensor $n$, consider the estimate sequence $\{\mathbf{x}_{n}(i)\}$ generated by the corresponding $\mathcal{GLU}$ algorithm with the above design parameters.
    Then, for every $0\leq\tau_{0}<\tau_{1}-\tau_{2}-\frac{1}{2+\varepsilon_{1}}$, we have
    \begin{equation}
    \label{main1:3}
    \mathbb{P}_{\mathbf{\theta}^{\ast}}\left(\lim_{i\rightarrow\infty}(i+1)^{\tau_{0}}\left(\mathbf{x}_{n}(i)-\mathbf{u}(i)\right)=0\right)=1,~~\forall n
    \end{equation}
\end{itemize}
\end{theorem}
We discuss the consequences of Theorem~\ref{main1}. The first assertion states that, as long as $0\leq\gamma_{0}<.5$, any distributed $\mathcal{GLU}$ estimator yields consistent parameter estimates at every sensor. By Remark~\ref{rem:prop-good}, this is precisely the class of fading parameters, a centralized estimator can estimate consistently. In other words, as long as a centralized linear estimator can consistently estimate a parameter, a distributed $\mathcal{GLU}$ estimator can. This is interesting, as the range of allowable $\gamma_{0}$s is independent of the network topology, and any random network satisfying the mean connectivity is sufficient. The second assertion quantifies the rate at which the distributed $\mathcal{GLU}$ estimator converges to the centralized estimator. Again, this rate is independent of the network topology.

The following result (Theorem~\ref{main2}) shows in what sense the $\mathcal{GLU}$ algorithm is optimal. We assume $\gamma_{0}=0$ in what follows. Suitable extensions to arbitrary $\gamma_{0}$ may be possible, however, this would impose added technicalities and digress from the main focus of the paper. Also, the notion of asymptotic variance as the metric for comparing different consistent estimators, is not quite clear for nonstationary recursive procedures.

\begin{theorem}
\label{main2}
\begin{itemize}
\item[(1)] Recall the positive definite matrix $G=\sum_{n=1}^{N}\overline{H}_{n}^{T}\overline{H}_{n}$. Assume $\tau_{1}=1$, i.e., the weight sequence $\{\alpha(i)\}$ is of the form
\begin{equation}
\label{uc3:3}
\alpha(i)=\frac{a}{i+1}
\end{equation}
where $a>\frac{N}{2\lambda_{\mbox{\scriptsize{min}}}(KG)}$ and $K$ is the positive definite matrix gain that commutes with $G$.
Choose any $\tau_{2}$ satisfying
\begin{equation}
\label{uc:th9}
\tau_{2}+\frac{1}{2+\varepsilon_{1}}<.5
\end{equation}
and note that such a choice exists as $\frac{1}{2+\varepsilon_{1}}<.5$. Consider the $\mathcal{GLU}$ algorithm with design parameters $(\tau_{1},a,\tau_{2},b,K)$ chosen above (this ensures that $(\tau_{1},a,\tau_{2},b,K)$ satisfy assumption (A.6).) Then, the
normalized estimate sequence $\{1/\sqrt(i+1)(\mathbf{x}_{n}(i)-\mathbf{\theta}^{\ast})\}$ is asymptotically normal for each $n$, i.e.,
\begin{equation}
\label{uc3:4}
\sqrt(i+1)\left(\mathbf{x}_{n}(i)-\mathbf{\theta}^{\ast}\right)\Longrightarrow\mathcal{N}(\mathbf{0},S_{c}(K))
\end{equation}
Here, the asymptotic variance $S_{c}(K)$ is the same obtained by a centralized estimator in Theorem~\ref{prop-centconv} with gain $K_{c}=K$.\\
\item[(2)] Let the hypothesis of the previous assertion hold with the matrix gain $K$ taking the value $K^{\ast}=G^{-1}$. Then, the asymptotic variance at each sensor is $S_{c}(K^{\ast})$, which is the asymptotic variance achieved by the best linear centralized estimator (see Proposition~\ref{prop-centconv}.) In particular, if the observation noise process is Gaussian, the $\mathcal{GLU}$ estimator constructed above is asymptotically efficient.
\end{itemize}
\end{theorem}
We interpret the above. The first assertion implies that given a centralized estimator with matrix gain $K$ and satisfying the assumptions in Proposition~\ref{prop-centconv}, there exists a distributed $\mathcal{GLU}$ estimator achieving the same asymptotic variance $S_{c}(K)$. This result is remarkable, as the asymptotic variance $S_{c}(K)$ is independent of the network topology $\overline{L}$. This is possible due to the mixed time scale behavior resulting from appropriate choice of $\tau_{1},\tau_{2}$. This invariance to the network topology is not achievable by the single time scale scheme ($\tau_{1}=\tau_{2}$) developed in~\cite{KarMouraRamanan-Est}. In a sense, Theorem~\ref{main2} justifies the applicability and advantage of distributed estimation schemes. Apart from issues of robustness, implementing a centralized estimator is much more communication intensive as it requires transmitting all sensor data to a fusion center at all times. On the other hand, the distributed $\mathcal{GLU}$ algorithm requires only sparse local communication among the sensors at each step, and achieves the performance of a centralized estimator asymptotically. The second assertion of the theorem reemphasizes the optimality and applicability of distributed estimation schemes, and shows that $\mathcal{GLU}$ can be designed to achieve the asymptotic variance of the optimal linear centralized scheme. In particular, if the observation noise process is Gaussian, $\mathcal{GLU}$ leads to asymptotically efficient estimators at each sensor.

\section{$\mathcal{GLU}$: Convergence properties}
\label{unq-conv}
As noted earlier, the mixed time scale behavior of $\mathcal{GLU}$ does not permit the use of standard stochastic approximation tools for establishing convergence. Moreover, to be able to establish important qualitative properties like asymptotic time scale separation, we need to clearly distinguish the long term effects of the consensus and innovations potential. We briefly outline the key steps involved in such a pursuit. We first identify conditions under which the sensor estimates $\{\mathbf{x}_{n}(i)\}$ converge to an \emph{averaged} estimate $\{\mathbf{x}_{\avg}(i)\}$ over the network and recognize the pathwise (strong) convergence rate. This is carried out in Lemma~\ref{uc-conv_l2}. The averaged estimator $\{\mathbf{x}_{\avg}(i)\}$ is not quite the centralized estimator $\{\mathbf{u}(i)\}$, the key reason being the averaged local innovations is not the centralized innovation. This leads us to study the rate of convergence of the averaged local innovations to the centralized innovation and hence, the convergence rate of the averaged estimate sequence to the centralized. This is accomplished in Lemma~\ref{uc4}. The analysis in all these steps culminate to Theorems~\ref{main1},\ref{main2}, the main results of the paper. These results identify conditions under which the consistent estimate sequences $\{\mathbf{x}_{n}(i)\}$ inherit the centralized convergence rate to $\mathbf{\theta}^{\ast}$. In particular, they establish sufficient conditions for the equivalence between the distributed and centralized schemes in terms of asymptotic variance. The methodology developed in this work is of independent interest and goes beyond the setting of distributed parameter estimation. We envision its applicability in the analysis of generic dynamical systems interacting over a network.

In what follows, we consider the $\mathcal{GLU}$ algorithm with fixed design parameters $(\tau_{1},a,\tau_{2},b,K)$ and assumptions (A.1)-(A.6) hold throughout.

We start by establishing pathwise boundedness of the sequence $\{\mathbf{x}(i)\}$.
\begin{lemma}
\label{unq-conv_l1}
There exists a finite random variable $R>0$, such that,
\begin{equation}
\label{unq-conv_l1:1}
\mathbb{P}_{\mathbf{\theta}^{\ast}}\left(\sup_{i\in\mathbb{T}_{+}}\left\|\mathbf{x}(i)\right\|\leq R\right)=1
\end{equation}
\end{lemma}
\begin{proof}
Define the process $\{\mathbf{y}(i)\}$ as
\begin{equation}
\label{uc1:2}
\mathbf{y}(i)=\mathbf{x}(i)-\mathbf{1}_{N}\otimes\mathbf{\theta}^{\ast}
\end{equation}
The assertion would follow if we establish boundedness for the process $\{\mathbf{y}(i)\}$. From eqn.~(\ref{unq-GLU2}) we note that $\{y(i)\}$ satisfies the recursion:
\begin{eqnarray}
\label{uc1:3}
\mathbf{y}(i+1) & = & \left(I_{NM}-\beta(i)\overline{L}\otimes I_{M}-\alpha(i)(I_{N}\otimes K)D_{\overline{H}}\right)\mathbf{y}(i)-\beta(i)\left(\widetilde{L}(i)\otimes I_{M}\right)\mathbf{y}(i)\nonumber \\ & & +\alpha(i)(I_{N}\otimes K)\left(\overline{D}_{\overline{H}}\mathbf{z}(i)-D_{\overline{H}}(\mathbf{1}_{N}\otimes\mathbf{\theta}^{\ast})\right)
\end{eqnarray}
where we use the invariance of the Laplacian operator,
\[
\left(\overline{L}\otimes I_{M}\right)\left(\mathbf{1}_{N}\otimes\mathbf{\theta}^{\ast}\right)=\mathbf{0}_{NM}
\]
Consider the process $\{V_{2}(i)\}$ given by
\begin{equation}
\label{uc1:4}
V_{2}(i)=\left\|\mathbf{y}(i)\right\|^{2}
\end{equation}
By using the conditional independence properties, it can be shown that,
\begin{eqnarray}
\label{uc1:5}
\mathbb{E}_{\mathbf{\theta}^{\ast}}\left[V_{2}(i+1)~|~\mathcal{F}_{i}\right] & = & V(i)+\beta^{2}(i)\mathbf{y}(i)^{T}\mathbb{E}_{\mathbf{\theta}^{\ast}}\left[\widetilde{L}^{2}(i)\right]\mathbf{y}(i)+\alpha^{2}(i)\mathbb{E}_{\mathbf{\theta}^{\ast}}\left[\left\|\overline{D}_{\overline{H}}\mathbf{z}(i)-D_{\overline{H}}(\mathbf{1}_{N}\otimes\mathbf{\theta}^{\ast})\right\|^{2}\right]\nonumber \\ & & -2\mathbf{y}^{T}(i)\left(\beta(i)\overline{L}\otimes I_{M}+\alpha(i)\left(I_{N}\otimes K\right)D_{\overline{H}}\right)\mathbf{y}(i)+\beta^{i}\mathbf{y}^{T}(i)(\overline{L}\otimes I_{M})^{2}\mathbf{y}(i)\nonumber \\ & & +\alpha^{2}(i)\mathbf{y}^{T}(i)\left(\left(I_{N}\otimes K\right)D_{\overline{H}}\right)^{T}\left(\left(I_{N}\otimes K\right)D_{\overline{H}}\right)\mathbf{y}(i)\nonumber \\ & & +2\alpha(i)\beta(i)\mathbf{y}^{T}(i)(\overline{L}\otimes I_{M})(I_{N}\otimes K)\mathbf{y}(i)
\end{eqnarray}
We use the following inequalities:
\begin{eqnarray}
\label{uc1:6}
\mathbf{y}(i)^{T}\mathbb{E}_{\mathbf{\theta}^{\ast}}\left[\widetilde{L}^{2}(i)\right]\mathbf{y}(i) & = & \mathbf{y}_{\mathcal{C}^{\perp}}^{T}(i)\mathbb{E}_{\mathbf{\theta}^{\ast}}\left[\widetilde{L}^{2}(i)\right]\mathbf{y}_{\mathcal{C}^{\perp}}(i)\nonumber \\ & \leq & c_{5}\left\|\mathbf{y}_{\mathcal{C}^{\perp}}(i)\right\|^{2}\\
\mathbf{y}^{T}(i)(\overline{L}\otimes I_{M})^{2}\mathbf{y}(i) & = & \mathbf{y}_{\mathcal{C}^{\perp}}^{T}(i)(\overline{L}\otimes I_{M})^{2}\mathbf{y}_{\mathcal{C}^{\perp}}(i)\nonumber \\ & \leq & \lambda_{N}^{2}(\overline{L})\left\|\mathbf{y}_{\mathcal{C}^{\perp}}(i)\right\|^{2}\\
2\mathbf{y}^{T}(i)\left(\beta(i)\overline{L}\otimes I_{M}+\alpha(i)\left(I_{N}\otimes K\right)D_{\overline{H}}\right)\mathbf{y}(i) & \geq & \beta(i)\mathbf{y}^{T}(i)(\overline{L}\otimes I_{M})\mathbf{y}(i)+\mathbf{y}^{T}(i)\left(\beta(i)\overline{L}\otimes I_{M}\right.\nonumber \\ & & \left.+\alpha(i)\left(I_{N}\otimes K\right)D_{\overline{H}}\right)\mathbf{y}(i)\nonumber \\ & \geq & \beta(i)\lambda_{2}(\overline{L})\left\|\mathbf{y}_{\mathcal{C}^{\perp}}(i)\right\|^{2}+c_{4}\alpha(i)\left\|\mathbf{y}(i)\right\|^{2}
\end{eqnarray}
We use Lemma~\ref{unq-GLU_lm1} to obtain the last inequality. Introducing additional constants to bound the quadratic forms and the moments, we derive the following from eqn.~(\ref{uc1:5}):
\begin{eqnarray}
\label{uc1:7}
\mathbb{E}_{\mathbf{\theta}^{\ast}}\left[V_{2}(i+1)~|~\mathcal{F}_{i}\right] & \leq & V_{2}(i)-\left(\beta(i)\lambda_{2}(\overline{L})-\beta^{2}(i)c_{5}-\beta^{2}(i)\lambda_{N}^{2}(\overline{L})\right)\left\|\mathbf{y}_{\mathcal{C}^{\perp}}(i)\right\|^{2}\nonumber \\ & & -\left(c_{4}\alpha(i)-\alpha(i)\beta(i)c_{7}\right)\left\|\mathbf{y}(i)\right\|^{2}+\alpha^{2}(i)\gamma^{2}(i)c_{8}+\alpha^{2}(i)c_{6}\left\|\mathbf{y}(i)\right\|^{2}
\end{eqnarray}
where $c_{8}>0$ is a constant, such that,
\begin{equation}
\label{uc1:189}
\alpha^{2}(i)\mathbb{E}_{\mathbf{\theta}^{\ast}}\left[\left\|\overline{D}_{\overline{H}}\mathbf{z}(i)-D_{\overline{H}}(\mathbf{1}_{N}\otimes\mathbf{\theta}^{\ast})\right\|^{2}\right] = \alpha^{2}(i)\gamma^{2}(i)c_{8}
\end{equation}
Since $\beta^{2}(i)$ goes to zero faster than $\beta(i)$, the $\beta(i)$ term dominates in the second expression of eqn.~(\ref{uc1:7}) eventually. Similarly, the $\alpha(i)$ term dominates the third expression eventually. Choose $c_{9}=\max(c_{6},c_{8})$. Since, $\gamma(i)\geq 1$ (assumption (A.2)), there exists $i_{2}$ large enough, such that, for $i\geq i_{2}$
\begin{eqnarray}
\label{uc1:8}
\mathbb{E}_{\mathbf{\theta}^{\ast}}\left[V_{2}(i+1)~|~\mathcal{F}_{i}\right]-V_{2}(i) & \leq & \alpha^{2}(i)\gamma^{2}(i)c_{8}+\alpha^{2}(i)c_{6}V_{2}(i)\nonumber \\ & \leq & c_{9}\alpha^{2}(i)\gamma^{2}(i)(1+V_{2}(i))
\end{eqnarray}
Now introduce the process
\begin{equation}
\label{uc1:9}
\widetilde{V}_{2}(i)=\left(1+V_{2}(i)\right)\prod_{k=i}^{\infty}(1+c_{9}\alpha^{2}(k)\gamma^{2}(k))
\end{equation}
Note that the above is well defined as the product $\prod_{k=i}^{\infty}(1+c_{9}\alpha^{2}(k)\gamma^{2}(k))$ converges for all $i$ due to the square summability of $\{\alpha(i)\gamma(i)\}$ (assumption (A.6)). Eqn.~(\ref{uc1:8}) and some algebraic manipulations lead to
\begin{equation}
\label{uc1:10}
\mathbb{E}_{\mathbf{\theta}^{\ast}}\left[\widetilde{V}_{2}(i+1)~|~\mathcal{F}_{i}\right]\leq \widetilde{V}_{2}(i)
\end{equation}
thus establishing that the sequence $\{\widetilde{V}_{2}(i)\}$ is a nonnegative supermartingale. Hence, there exists a finite random variable $\widetilde{R}$, such that, $\lim_{i\rightarrow\infty}\widetilde{V}_{2}(i)=R$ a.s. We then have from eqn.~(\ref{uc1:9})
\begin{equation}
\label{uc1:11}
\lim_{i\rightarrow\infty}V_{2}(i)=\widetilde{R}-1~\mbox{a.s.}
\end{equation}
Hence, $\{V_{2}(i)\}$ is bonded pathwise and the assertion follows.
\end{proof}
\begin{remark} A deeper investigation of the supermartingale would reveal that $V_{2}(i)$ in fact, converges to zero. This would have established the consistency of the estimators.
However, to obtain strong convergence rates, we need to study the sample paths more critically. The rest of this subsection is devoted to this study.
\end{remark}

The following lemma identifies the rate at which the estimates converge to a network \emph{averaged estimate} and hence characterizes the information flow in the network.

Before that, we establish the following:
\begin{proposition}
\label{prop-inter}
Let assumptions (A.1)-(A.6) hold.
\begin{itemize}
\item[(1)]
For all $i\in\mathbb{T}_{+}$, define
\begin{equation}
\label{uc2:8}
J_{1}(\mathbf{z}(i))=(I_{N}\otimes K)\overline{D}_{\overline{H}}\mathbf{z}(i)-\mathbf{1_{N}}\otimes\left(\left(\frac{1}{N}\mathbf{1}_{N}\otimes I_{M}\right)(I_{N}\otimes K)\overline{D}_{\overline{H}}\mathbf{z}(i)\right)
\end{equation}
Then, we have the following:
\begin{equation}
\label{uc2:1090}
\mathbb{P}_{\mathbf{\theta}^{\ast}}\left(\frac{1}{(i+1)^{\gamma_{0}+\frac{1}{2+\varepsilon_{1}}+\delta}}\left\|J_{1}(\mathbf{z}(i))\right\|=0\right)=0
\end{equation}
\item[(2)] Recall the matrix,
\begin{equation}
\label{uc2:4} P^{NM}=\frac{1}{N}\left(\mathbf{1}_{N}\otimes
I_{M}\right)\left(\mathbf{1}_{N}\otimes I_{M}\right)^{T}
\end{equation}
Then, for $i\in\mathbb{T}_{+}$ sufficiently large, we have
\begin{equation}
\label{prop-inter1}
\left\|I_{NM}-\beta(i)\left(L(i)\otimes I_{M}\right)-P^{NM}\right\|=1-\beta(i)\lambda_{2}(L(i))
\end{equation}
\end{itemize}
\end{proposition}
\begin{proof}
For the first assertion, consider any $\varepsilon_{2}>0$. By Chebyshev's inequality and assumption (A.5),
\begin{eqnarray}
\label{prop-inter2}
\mathbb{P}_{\mathbf{\theta}^{\ast}}\left(\frac{1}{(i+1)^{\frac{1}{2+\varepsilon_{1}}+\delta}}\left\|J_{1}(\mathbf{z}(i))\right\|>\varepsilon_{2}\right) & \leq & \frac{1}{\varepsilon_{2}^{2+\varepsilon_{1}}(i+1)^{1+(\delta+\gamma_{0})(2+\varepsilon_{1})}}\mathbb{E}_{\mathbf{\theta}^{\ast}}\left[\left\|J_{1}(\mathbf{z}(i))\right\|^{2+\varepsilon_{1}}\right]
\nonumber \\ & = & \frac{\kappa(\mathbf{\theta}^{\ast})}{\varepsilon_{2}^{2+\varepsilon_{1}}}\frac{1}{(i+1)^{1+\delta(2+\varepsilon_{1})}}
\end{eqnarray}
Since, $\delta>0$, the sequence $\{\frac{1}{(i+1)^{1+\delta(2+\varepsilon_{1})}}\}$ is square summable and we obtain
\begin{equation}
\label{prop-inter3}
\sum_{i\in\mathbb{T}_{+}}\mathbb{P}_{\mathbf{\theta}^{\ast}}\left(\frac{1}{(i+1)^{\frac{1}{2+\varepsilon_{1}}+\delta}}\left\|J_{1}(\mathbf{z}(i))\right\|>\varepsilon_{2}\right)<\infty
\end{equation}
It then follows from the Borel-Cantelli lemma (see~\cite{Kallenberg}) that,
\begin{equation}
\label{uc2:12}
\mathbb{P}_{\mathbf{\theta}^{\ast}}\left(\frac{1}{(i+1)^{\frac{1}{2+\varepsilon_{1}}+\delta}}\left\|J_{1}(\mathbf{z}(i))\right\|>\varepsilon_{2}~\mbox{i.o.}\right)=0
\end{equation}
where i.o. stands for infinitely often. Since the above holds for $\varepsilon_{2}>0$ arbitrarily small, the claim in eqn.~(\ref{uc2:1090}) holds by standard arguments.

For the second assertion, we note from the discussion on Kronecker products in Section~\ref{notation} that, the eigenvalues of the matrix $\left(I_{NM}-\beta(i)\left(L(i)\otimes I_{M}\right)-P^{NM}\right)$ are 0 and $1-\beta(i)\lambda_{n}(L(i))$, $i=2,\cdots,N$, each repeated $M$ times. Since, the Laplacian eigenvalues are all bounded above by $N^{2}$ and $\beta(i)\rightarrow 0$, there exists $i_{4}\in\mathbb{T}_{+}$ sufficiently large, such that, for $i\geq i_{4}$, $\beta(i)\lambda_{n}(L(i))< 1$, for all $2\leq n\leq N$. The assertion is then obvious.
\end{proof}

\begin{lemma}
\label{uc-conv_l2}
Define the averaged estimate sequence $\{\mathbf{x}_{\avg}(i)\}$ as
\begin{equation}
\label{uc2:1}
\mathbf{x}_{\avg}(i)=\frac{1}{N}(\mathbf{1}_{N}\otimes I_{M})\mathbf{x}(i)
\end{equation}
Then for every $\tau_{0}$, such that,
\begin{equation}
\label{uc2:2}
0\leq \tau_{0}<\tau_{1}-\tau_{2}-\gamma_{0}-\frac{1}{2+\varepsilon}
\end{equation}
we have
\begin{equation}
\label{uc2:3}
\mathbb{P}_{\mathbf{\theta}^{\ast}}\left(\lim_{i\rightarrow\infty}(i+1)^{\tau_{0}}\left(\mathbf{x}(i)-\mathbf{1}_{N}\otimes\mathbf{x}_{\avg}(i)\right)=0\right)=1
\end{equation}
\end{lemma}
\begin{proof}
Define the process $\{\mathbf{y}_{1}(i)\}$:
\begin{equation}
\label{uc2:6}
\widehat{\mathbf{y}}(i)=\mathbf{x}(i)-\mathbf{1}_{N}\otimes\mathbf{x}_{\avg}(i)
\end{equation}
Recall the matrix
\begin{equation}
\label{uc2:499} P^{NM}=\frac{1}{N}\left(\mathbf{1}_{N}\otimes
I_{M}\right)\left(\mathbf{1}_{N}\otimes I_{M}\right)^{T}
\end{equation}
and note that
\begin{equation}
\label{uc2:5}
P^{NM}\mathbf{x}(i)=\mathbf{1}_{N}\otimes\mathbf{x}_{\avg}(i),~~P^{NM}\left(\mathbf{1}_{N}\otimes\mathbf{x}_{\avg}(i)\right)=\mathbf{1}_{N}\otimes\mathbf{x}_{\avg}(i)
\end{equation}
From eqn.~(\ref{unq-GLU2}) we then note that $\{\mathbf{y}_{1}(i)\}$ satisfies the recursion:
\begin{eqnarray}
\label{uc2:7}
\widehat{\mathbf{y}}(i+1)&=&\left(I_{NM}-\beta(i)\overline{L}\otimes I_{M}-P^{NM}\right)\widehat{\mathbf{y}}(i)-\alpha(i)\left[(I_{N}\otimes K)D_{\overline{H}}\mathbf{x}(i)\right.\nonumber \\ & & \left. -\mathbf{1}_{N}\otimes\left(\frac{1}{N}(\mathbf{1}_{N}\otimes I_{M})(I_{N}\otimes K)D_{\overline{H}}\mathbf{x}(i)\right)\right]\nonumber \\ & & +\alpha(i)\left[J_{1}(\mathbf{z}(i))\right]
\end{eqnarray}
where $J_{1}(\mathbf{z}(i))$ is defined in~\eqref{uc2:8}.
Choose $\delta$ satisfying
\begin{equation}
\label{uc2:9}
0<\delta<\tau_{1}-\tau_{2}-\gamma_{0}-\tau_{0}-\frac{1}{2+\varepsilon_{1}}
\end{equation}
Then, by Proposition~\ref{prop-inter}, we have
\begin{equation}
\label{uc2:10}
\mathbb{P}_{\mathbf{\theta}^{\ast}}\left(\frac{1}{(i+1)^{\gamma_{0}+\frac{1}{2+\varepsilon_{1}}+\delta}}\left\|J_{1}(\mathbf{z}(i))\right\|=0\right)=0
\end{equation}
Also, Lemma~\ref{unq-conv_l1} implies
\begin{equation}
\label{uc2:13}
\mathbb{P}_{\mathbf{\theta}^{\ast}}\left(\sup_{i\in\mathbb{T}_{+}}\left\|(I_{N}\otimes K)D_{\overline{H}}\mathbf{x}(i)-\mathbf{1}_{N}\otimes\left(\frac{1}{N}(\mathbf{1}_{N}\otimes I_{M})(I_{N}\otimes K)D_{\overline{H}}\mathbf{x}(i)\right)\right\|<\infty\right)=1
\end{equation}
by the boundedness of $\{\mathbf{x}(i)\}$.
However, these pathwise bounds are not uniform over the sample paths and hence we use truncation arguments. For a
scalar $a$, define its truncation $(a)^{R_{0}}$ at level $R_{0}>0$ by
\begin{equation}
\label{uc2:14} (a)^{R_{0}}=\left\{\begin{array}{ll}
                                \frac{a}{|a|}\min(|a|,R_{0}) & \mbox{if
                                $a\neq 0$}\\
                                0 & \mbox{if $a=0$}
                                \end{array}
                                \right.
\end{equation}
For a vector, the truncation operation applies component-wise. For
$R_{0}>0$, we also consider the sequences,
$\left\{\widehat{\mathbf{y}}_{R_{0}}(i)\right\}_{i\geq 0}$, given by
\begin{eqnarray}
\label{uc2:15}
\widehat{\mathbf{y}}_{R_{0}}(i+1) & =& \left(I_{NM}-\beta(i)\overline{L}\otimes I_{M}-P\right)\widehat{\mathbf{y}}_{R_{0}}(i)-\alpha(i)\left(\left[(I_{N}\otimes K)D_{\overline{H}}\mathbf{x}(i)-\right.\right. \nonumber \\ & & \left.\left.\mathbf{1}_{N}\otimes\left(\frac{1}{N}(\mathbf{1}_{N}\otimes I_{M})(I_{N}\otimes K)D_{\overline{H}}\mathbf{x}(i)\right)\right]\right)^{R_{0}}\nonumber \\ & & +\alpha(i)\left(\left[J_{1}(\mathbf{z}(i))\right]\right)^{R_{0}(i+1)^{\gamma_{0}+\frac{1}{2+\varepsilon_{1}}+\delta}}
\end{eqnarray}
We will now show that for every $R_{0}>0$,
\begin{equation}
\label{uc2:16}
\mathbb{P}_{\mathbf{\theta}^{\ast}}\left(\lim_{i\rightarrow\infty}(i+1)^{\tau_{0}}\left(\widehat{\mathbf{y}}_{R_{0}}(i)\right)=0\right)=1
\end{equation}
for $\tau_{0}$ satisfying the hypothesis~\ref{uc2:2}. That, this is sufficient to conclude the assertion
\begin{equation}
\label{uc2:17}
\mathbb{P}_{\mathbf{\theta}^{\ast}}\left(\lim_{i\rightarrow\infty}(i+1)^{\tau_{0}}\left(\widehat{\mathbf{y}}(i)\right)=0\right)=1
\end{equation}
is a consequence of the following standard argument. The pathwise boundedness of the various terms imply that for every $\varepsilon_{3}>0$, there exists $R_{\varepsilon_{3}}>0$, such that,
\begin{equation}
\label{uc2:18}
\mathbb{P}_{\mathbf{\theta}^{\ast}}\left(\sup_{i\in\mathbb{T}_{+}}\left\|(I_{N}\otimes K)D_{\overline{H}}\mathbf{x}(i)-\mathbf{1}_{N}\otimes\left(\frac{1}{N}(\mathbf{1}_{N}\otimes I_{M})(I_{N}\otimes K)D_{\overline{H}}\mathbf{x}(i)\right)\right\|<R_{\varepsilon_{3}}\right)>1-\varepsilon_{3}
\end{equation}
\begin{equation}
\label{uc2:19}
\mathbb{P}_{\mathbf{\theta}^{\ast}}\left(\sup_{i\in\mathbb{T}_{+}}\left\|J_{1}(\mathbf{z}(i))\right\|<R_{\varepsilon_{3}}(i+1)^{\gamma_{0}+\frac{1}{2+\varepsilon_{1}}+\delta}\right)>1-\varepsilon_{3}
\end{equation}
For~\eqref{uc2:18} we use the pathwise boundedness of $\{\mathbf{x}(i)\}$ (Lemma~\ref{unq-conv_l1}), whereas,~\eqref{uc2:19} holds because the a.s. convergence in Lemma~\ref{prop-inter} implies convergence in probability.
Clearly, the process $\{\widehat{\mathbf{y}}(i)\}$ agrees with the process $\{\widehat{\mathbf{y}}_{R_{\varepsilon_{3}}}(i)\}$ on the set where both of the above events occur. By standard manipulations, it then follows, that
\begin{equation}
\label{uc2:20}
\mathbb{P}_{\mathbf{\theta}^{\ast}}\left(\sup_{i\in\mathbb{T}_{+}}\left\|\widehat{\mathbf{y}}(i)-\widehat{\mathbf{y}}_{R_{\varepsilon_{3}}}(i)\right\|=0\right)>1-2\varepsilon_{3}
\end{equation}
The claim in eqn.~(\ref{uc2:16}) would then imply
\begin{equation}
\label{uc2:21}
\mathbb{P}_{\mathbf{\theta}^{\ast}}\left(\lim_{i\rightarrow\infty}(i+1)^{\tau_{0}}\left(\widehat{\mathbf{y}}(i)\right)=0\right)>1-2\varepsilon_{3}
\end{equation}
We could then establish the assertion of the lemma by taking $\varepsilon_{3}$ to zero.

Hence, in the following we establish the claim in eqn.~(\ref{uc2:16}) for every $R_{0}>0$. To this end, consider the scalar process $\{\widetilde{y}_{R_{0}}(i)\}_{i\in\mathbb{T}_{+}}$ defined recursively as
\begin{equation}
\label{uc2:22}
\widetilde{y}_{R_{0}}(i+1)=\left\|I_{NM}-\beta(i)L(i)-P^{NM}\right\|\widetilde{y}_{R_{0}}(i)+NMR_{0}\alpha(i)+NMR_{0}\alpha(i)(i+1)^{\gamma_{0}+\frac{1}{2+\varepsilon_{1}}+\delta}
\end{equation}
with initial condition $\widetilde{y}_{R_{0}}(0)=\|\widehat{\mathbf{y}}_{R_{0}}(0)\|$. Since,
\begin{eqnarray}
\label{uc2:23}
\left\|\widehat{\mathbf{y}}_{R_{0}}(i+1)\right\| & = & \left\|I_{NM}-\beta(i)\overline{L}\otimes I_{M}-P^{NM}\right\|\left\|\widehat{\mathbf{y}}_{R_{0}}(i)\right\|-\alpha(i)\left\|\left(\left[(I_{N}\otimes K)D_{\overline{H}}\mathbf{x}(i)\right.\right.\right.\nonumber \\ & & \left.\left.\left.-\mathbf{1}_{N}\otimes\left(\frac{1}{N}(\mathbf{1}_{N}\otimes I_{M})(I_{N}\otimes K)D_{\overline{H}}\mathbf{x}(i)\right)\right]\right)^{R_{0}}\right\|\nonumber \\ & & +\alpha(i)\left\|\left(\left[J_{1}(\mathbf{z}(i))\right]\right)^{R_{0}(i+1)^{\gamma_{0}+\frac{1}{2+\varepsilon_{1}}+\delta}}\right\|
\end{eqnarray}
it follows that,
\begin{equation}
\label{uc2:24}
\left\|\widehat{\mathbf{y}}_{R_{0}}(i)\right\|\leq\widetilde{y}_{R_{0}}(i),~~\forall i
\end{equation}
By Proposition~\ref{prop-inter}, for $i$ large enough, it can be shown that
\begin{equation}
\label{uc2:25}
\left\|I_{NM}-\beta(i)\overline{L}\otimes I_{M}-P^{NM}\right\|=1-\beta(i)\lambda_{2}(L(i))
\end{equation}
We assume w.l.o.g. that the above holds for all $i$. We then have
\begin{eqnarray}
\label{uc2:26}
\widetilde{y}_{R_{0}}(i+1)& \leq & \left(1-\beta(i)\lambda_{2}(L(i))\right)\widetilde{y}_{R_{0}}(i)+NMR_{0}\alpha(i)+NMR_{0}\alpha(i)(i+1)^{\gamma_{0}+\frac{1}{2+\varepsilon_{1}}+\delta}\nonumber \\ & \leq &
\left(1-\beta(i)\lambda_{2}(L(i))\right)\widetilde{y}_{R_{0}}(i)+2NMR_{0}\alpha(i)(i+1)^{\gamma_{0}+\frac{1}{2+\varepsilon_{1}}+\delta}
\end{eqnarray}
The above implies
\begin{equation}
\label{uc2:27}
\widetilde{y}_{R_{0}}(i+1)\leq\left(1-\beta(i)\lambda_{2}(L(i))\right)\left(\widetilde{y}_{R_{0}}(i)\right)+2NMR_{0}\frac{1}{(i+1)^{\tau_{1}-\gamma_{0}-\frac{1}{2+\varepsilon_{1}}-\delta}}
\end{equation}
Using a result from~\cite{tsp07-K-M}, we note that $\lambda_{2}(\overline{L})>0$ implies $\mathbb{E}_{\mathbf{\theta}^{\ast}}\left[\lambda_{2}(L(i))\right]>0$ (note that this equivalence is not a consequence of Jensen's inequality, as the second eigenvalue is a concave function of the graph Laplacian.) The recursion in eqn.~(\ref{uc2:28}) then falls under the purview of Lemmas~\ref{unq-GLU_lemma},\ref{unq-prop} (see eqns.~(\ref{uc2:2},\ref{uc2:9})), and we have
\begin{equation}
\label{uc2:28}
\mathbb{P}_{\mathbf{\theta}^{\ast}}\left(\lim_{i\rightarrow\infty}(i+1)^{\tau_{0}}\widetilde{y}_{R_{0}}(i)=0\right)=1
\end{equation}
It then follows from eqn.~(\ref{uc2:24}) that
\begin{equation}
\label{uc2:29}
\mathbb{P}_{\mathbf{\theta}^{\ast}}\left(\lim_{i\rightarrow\infty}(i+1)^{\tau_{0}}\widehat{\mathbf{y}}_{R_{0}}(i)=0\right)=1
\end{equation}
The assertion is then immediate.
\end{proof}

Lemma~\ref{uc-conv_l2} characterizes the proximity of the sensor estimates $\{\mathbf{x}_{n}(i)\}$ to the network averaged estimate $\{\mathbf{x}_{\avg}(i)\}$. To infer the convergence of the sensor estimates to $\mathbf{\theta}^{\ast}$, it then suffices to study the limiting properties of $\{\mathbf{x}_{\avg}(i)\}$. This is achieved in two steps. In the following, we consider the class of linear centralized estimators of the parameter $\theta$, and establish its relation to the network averaged estimator $\{\mathbf{x}_{\avg}(i)\}$. In particular, we investigate the rate at which $\{\mathbf{x}_{\avg}(i)\}$ converges to the class of centralized estimators. Properties of the centralized estimators are then used to infer the convergence of $\{\mathbf{x}_{\avg}(i)\}$ (and hence, that of $\{\mathbf{x}_{n}(i)\}$) to $\mathbf{\theta}^{\ast}$.

The following result is the first step towards characterizing the convergence rate of the network averaged estimator $\{\mathbf{x}_{\avg}(i)\}$ to $\mathbf{\theta}^{\ast}$. It establishes the relation between $\{\mathbf{x}_{\avg}(i)\}$ and the class of centralized estimators $\{\mathbf{u}(i)\}$ introduced in Definition~\ref{def_centest}.

\begin{lemma}
\label{uc4}Let $\{\mathbf{u}(i)\}$ be the centralized estimate sequence defined in~\ref{def_centest} with $\tau_{c}=\tau_{1}$, $a_{c}=a$ and $K_{c}=K$. Then,
\begin{itemize}
\item[(1)]
\begin{equation}
\label{uc4:786}
\mathbb{P}_{\mathbf{\theta}^{\ast}}\left(\lim_{i\rightarrow\infty}\left\|\mathbf{x}_{\avg}(i)-\mathbf{u}(i)\right\|=0\right)=1
\end{equation}
\item[(2)]
Let $\tau_{0}$ satisfy the assumption
\begin{equation}
\label{uc4:1}0<\tau_{0}<\tau_{1}-\tau_{2}-\gamma_{0}-\frac{1}{2+\varepsilon_{1}}
\end{equation}
Also, if $\tau_{1}=1$, assume that the constant $a$ in assumption~\textbf{(A.6)} satisfies
\begin{equation}
\label{uc4:1000}
a>\frac{N\tau_{0}}{\lambda_{\mbox{\scriptsize{min}}}(KG)}
\end{equation}
Then,
\begin{equation}
\label{uc4:2}
\lim_{i\rightarrow\infty}(i+1)^{\tau_{0}}\left(\mathbf{x}_{\avg}(i)-\mathbf{u}(i)\right)=0
\end{equation}

\end{itemize}
\end{lemma}
\begin{proof} We note that the averaged update may be written as
\begin{eqnarray}
\label{uc4:4}
\mathbf{x}_{\avg}(i+1) & =& \mathbf{x}_{\avg}(i)+\frac{\alpha(i)}{N}K\sum_{n=1}^{N}\overline{H}_{n}^{T}\mathbf{z}_{n}(i)-\frac{\alpha(i)}{N}K\sum_{n=1}^{N}\overline{H}_{n}^{T}\overline{H}_{n}\mathbf{x}_{n}(i)\nonumber \\ & = &
\mathbf{x}_{\avg}(i)+\frac{\alpha(i)}{N}K\sum_{n=1}^{N}\left(\overline{H}_{n}^{T}\mathbf{z}_{n}(i)-\overline{H}_{n}^{T}\overline{H}_{n}\mathbf{x}_{\avg}(i)\right)\nonumber \\ & & -\frac{\alpha(i)}{N}K\sum_{n=1}^{N}\overline{H}_{n}^{T}\overline{H}_{n}\left(\mathbf{x}_{n}(i)-\mathbf{x}_{\avg}(i)\right)
\end{eqnarray}
Define the process $\{\widetilde{\mathbf{u}}(i)\}$ by
\begin{equation}
\label{uc4:5}
\widetilde{\mathbf{u}}(i)=\mathbf{x}_{\avg}(i)-\mathbf{u}(i)
\end{equation}
We then have
\begin{equation}
\label{uc4:6}
\widetilde{\mathbf{u}}(i+1)=(I_{M}-\frac{\alpha(i)}{N}KG)\widetilde{\mathbf{u}}(i)-\frac{\alpha(i)}{N}K\sum_{n=1}^{N}\overline{H}_{n}^{T}\overline{H}_{n}\left(\mathbf{x}_{n}(i)-\mathbf{x}_{\avg}(i)\right)
\end{equation}
Now choose $\delta$, such that,
\begin{equation}
\label{uc4:7}
0<\delta<\tau_{1}-\tau_{2}-\gamma_{0}-\tau_{0}-\frac{1}{2+\varepsilon_{1}}
\end{equation}
Since $\tau_{0}+\delta<\tau_{1}-\tau_{2}-\gamma_{0}-\frac{1}{2+\varepsilon_{1}}$, by Lemma~\ref{uc-conv_l2}, it follows that,
\begin{equation}
\label{uc4:890}
\mathbb{P}_{\mathbf{\theta}^{\ast}}\left(\lim_{i\rightarrow\infty}(i+1)^{\tau_{0}+\delta}\left\|\sum_{n=1}^{N}\overline{H}_{n}^{T}\overline{H}_{n}\left(\mathbf{x}_{n}(i)-\mathbf{x}_{\avg}(i)\right)\right\|=0\right)=1
\end{equation}
Then, there exists a finite random variable $R_{3}$, such that,
\begin{equation}
\label{uc4:8}
\mathbb{P}_{\mathbf{\theta}^{\ast}}\left(\left\|\sum_{n=1}^{N}\overline{H}_{n}^{T}\overline{H}_{n}\left(\mathbf{x}_{n}(i)-\mathbf{x}_{\avg}(i)\right)\right\|\leq R_{3}(i+1)^{-\tau_{0}-\delta}~~\forall i\in\mathbb{T}_{+}\right)=1
\end{equation}
Note, by hypothesis, the matrix $KG$ is symmetric and $\alpha(i)\rightarrow 0$. Hence, there exists a constant $c_{10}>0$, such that, for sufficiently large $i$,
\[
\left\|I_{M}-\frac{\alpha(i)}{N}KG\right\|\leq 1-c_{10}\alpha(i)
\]
Writing $\omega$-wise and introducing another constant $c_{11}>0$, we have
\begin{equation}
\label{uc4:9}
\left\|\widetilde{\mathbf{u}}(i+1,\omega)\right\|\leq (1-c_{10}\alpha(i))\left\|\widetilde{\mathbf{u}}(i,\omega)\right\|+c_{11}\alpha(i)R_{3}(\omega)(i+1)^{-\tau_{0}-\delta}
\end{equation}
for $i$ greater than some sufficiently large $i_{4}(\omega)$. We then have
\begin{equation}
\label{uc4:10}
\left\|\widetilde{\mathbf{u}}(i+1,\omega)\right\|\leq (1-c_{KG}\alpha(i))\left\|\widetilde{\mathbf{u}}(i,\omega)\right\|+c_{11}R_{3}(\omega)(i+1)^{-\tau_{1}-\tau_{0}-\delta}
\end{equation}
A pathwise (fixed $\omega$) application of Lemma~\ref{unq-GLU_lemma} and Lemma~\ref{unq-prop} and noting that the above holds for $\omega$ in a set of full measure yield the assertions.
\end{proof}

\section{Proofs of main results}
\label{proof_main_res}

\subsection{Proof of Theorem~\ref{main1}}
\label{proof_1}
Consider the first assertion. Since the $\mathcal{GLU}$ parameters $(\tau_{1},a,\tau_{2},b,K)$ satisfy assumption (A.6), we note
\begin{equation}
\label{proof_1:1}
.5+\gamma_{0}<\tau_{1}\leq 1
\end{equation}
Choose
$\tau_{c}=\tau_{1}$ and $K_{c}=K$. It then follows that the centralized estimator $\{\mathbf{u}(i)\}$ (Definition~\ref{def_centest}) with design parameters is good. Hence, by Proposition~\ref{prop-good} it is consistent, i.e.,
\begin{equation}
\label{proof_1:2}
\mathbb{P}_{\mathbf{\theta}^{\ast}}\left(\lim_{i\rightarrow\infty}\mathbf{u}(i)=\mathbf{\theta}^{\ast}\right)=1
\end{equation}
Taking $\tau_{0}=0$ in Lemma~\ref{uc-conv_l2}, we have
\begin{equation}
\label{proof_1:3}
\mathbb{P}_{\mathbf{\theta}^{\ast}}\left(\lim_{i\rightarrow\infty}\left(\mathbf{x}(i)-\mathbf{1}_{N}\otimes\mathbf{x}_{\avg}(i)\right)=0\right)=1
\end{equation}
The first assertion of Theorem~\ref{main1} is then an immediate consequence of~\eqref{proof_1:2}-\eqref{proof_1:3} and Lemma~\ref{uc4} (first assertion.)

The second assertion of Theorem~\ref{main1} is a direct consequence of Lemma~\ref{uc-conv_l2} and Lemma~\ref{uc4} (first assertion.)

\subsection{Proof of Theorem~\ref{main2}}
\label{proof_2}
By hypothesis of Theorem~\ref{main2}, we have
\begin{equation}
\label{proof_2:1}
\tau_{1}=1,~~\frac{1}{2+\varepsilon_{1}}+\tau_{2}<.5
\end{equation}
Hence, $\tau_{1}-\tau_{2}-\frac{1}{2+\varepsilon_{1}}>.5$.
Since, $a>\frac{N}{2\lambda_{\mbox{\scriptsize{min}}}(KG)}$, there exists $\varepsilon_{5}>0$, small enough, such that,
\begin{equation}
\label{proof_2:2}
a>\frac{N(.5+\varepsilon_{5})}{\lambda_{\mbox{\scriptsize{min}}}(KG)}
\end{equation}
By the above, we can always choose $\tau_{0}$ satisfying the condition:
\begin{equation}
\label{proof_2:3}
.5<\tau_{0}<\max\left(.5+\varepsilon_{5},\tau_{1}-\tau_{2}-\frac{1}{2+\varepsilon_{1}}\right)
\end{equation}
For such $\tau_{0}$, we clearly have $a>\frac{N\tau_{0}}{\lambda_{\mbox{\scriptsize{min}}}(KG)}$, and hence by Theorem~\ref{main1} (second assertion), we conclude
\begin{equation}
\label{proof_2:4}
\mathbb{P}_{\mathbf{\theta}^{\ast}}\left(\lim_{i\rightarrow\infty}(i+1)^{\tau_{0}}\left(\mathbf{x}_{n}(i)-\mathbf{u}(i)\right)=0\right)=1
\end{equation}
where $\{\mathbf{u}(i)\}$ is the centralized estimator with design parameters $(\tau_{c},a_{c},K_{c})$, such that, $a_{c}=a$, $\tau_{c}=\tau_{1}$, $K_{c}=K$. It then follows by Proposition~\ref{prop-centconv}, that,
\begin{equation}
\label{proof_2:5}
\sqrt(i+1)\left(\mathbf{u}(i)-\mathbf{\theta}^{\ast}\right)\Longrightarrow\mathcal{N}(\mathbf{0},S_{c}(K))
\end{equation}
Since, $\tau_{0}$ in~\eqref{proof_2:4} is strictly greater than .5, the sequences $\{\mathbf{x}_{n}(i)\}$ and $\{\mathbf{u}(i)\}$ are indistinguishable in $\sqrt(i+1)$ scale, and it can be shown using standard properties of stochastic convergence, that,
\begin{equation}
\label{proof_2:6}
\sqrt(i+1)\left(\mathbf{x}_{n}(i)-\mathbf{\theta}^{\ast}\right)\Longrightarrow\mathcal{N}(\mathbf{0},S_{c}(K))
\end{equation}

The second assertion follows by choosing $K=K^{\ast}$ in the first.

\section{Conclusions}
\label{conclusion}
The paper considers gossip linear estimation of an unknown large dimensional parameter (or large scale static random field) observed by a sparsely interconnected network of sensors operating under the gossip communication protocol. We consider this problem under very general conditions on the noise assumptions and communication failures (including, link or channel failures, besides the usual measurement noise assumptions.) Due to the large scale of the field, the sensors are local, i.e., they observe only a small fraction of the field. To obtain a global estimate, the sensors need to cooperate. The class of gossip distributed linear estimators we study combines two terms: a \emph{consensus} term that updates at each sensor its current estimate with the state estimates provided by the neighbor(s) when they gossip; and an \emph{innovations} or \emph{sensing} term that updates the current sensor estimate with the new observation. The linear gossip distributed estimators that we analyze exhibit a mixed time scale--one that is associated with the consensus and the other with the innovations. This forces us to develop new analytical tools to establish their asymptotic properties. This is because in gossip distributed estimation,  the innovation term is not a martingale difference process, as in previous work on mixed time scale stochastic approximation algorithms, e.g., \cite{Gelfand-Mitter}; so, a key step in our analysis is to derive pathwise strong approximation results to characterize the rate at which the innovation process converges to a martingale difference process. The paper establishes a distributed observability condition--global observability, a condition on the sensing devices, i.e., the local measurements, plus mean connectedness, a structural condition on the communication network as provided by gossip. We show that under this condition the distributed estimators performance approaches the asymptotic performance of the optimal centralized estimators, namely, the distributed estimators are consistent and asymptotically normal. This is significant, as it shows that, under reasonable assumptions, a distributed gossip estimator is as good as a centralized one, the latter having access to all sensor observations at all times. As mentioned, the distributed gossip estimator has two time scales, which involves setting two gain sequences, one for the local innovations at each sensor and the other for estimate fusion (consensus) across sensors. To design good distributed gossip estimators, these gains should be chosen properly, namely, the consensus gain should decay at a slower rate than the innovation gain. In the absence of quantization or channel noise, the paper shows that it is possible to choose the consensus weight sequence such that its squared sum goes to $\infty$, in contrast to the innovation weight sequence whose squared sum needs to be finite. This tuning of the different gain sequences leads to an asymptotic time scale separation, the rate of information dissemination dominating the rate of reduction of uncertainty by observation acquisition. This is not possible with quantized or noisy transmissions, as each consensus step introduces noise, preventing proper adjustment of the gain sequences. The paper interprets the fundamental convergence results on distributed gossip estimation in two interesting contexts: \begin{inparaenum}[1)] \item when the observations are (conditionally) independent, the distributed estimator achieves the same performance (in terms of asymptotic variance) as the best centralized linear estimator; and \item the maximum rate at which the observation noise power (variance) can increase with time and still the estimators to remain consistent is the same for the centralized and the gossip linear distributed estimators.
\end{inparaenum}

\bibliographystyle{IEEEtran}
\bibliography{IEEEabrv,CentralBib}

\end{document}